\documentclass[12pt, a4paper]{article}
\usepackage[utf8]{inputenc}

\usepackage[margin=1.25in]{geometry}
\usepackage{natbib}
\setcitestyle{authoryear,open={(},close={)}}
\usepackage{titlesec}
\titleformat*{\section}{\large\bfseries}
\titleformat*{\subsection}{\normalsize\bfseries}
\titleformat*{\subsubsection}{\normalsize\bfseries}
\usepackage{mathtools}

\usepackage{comment}
\usepackage{xcolor}
\usepackage{tikz}
\usetikzlibrary{patterns}
\usepackage{graphics}
\usepackage{graphicx}
\usepackage{color}
\usepackage{amsmath}
\usepackage{amsthm}
\usepackage{amsfonts}
\usepackage[normalem]{ulem}
\usepackage{float}
\usepackage{dsfont} % for \mathds{1}

\usepackage{mathabx} %% for \updownarrows 

\usepackage[breaklinks]{hyperref}
\hypersetup{colorlinks=true,linkcolor=blue,citecolor=blue}
\usepackage{nicefrac}

\ifdefined\DRAFT

\definecolor{ForestGreen}{rgb}{.13,.54,.13}
\definecolor{violet}{cmyk}{0.79,0.88,0,0}
\newcommand{\fed}[1]{{\color{ForestGreen}{(\textbf{Fedor:} #1)}}}
\newcommand{\omer}[1]{{\color{blue}{(\textbf{Omer:} #1)}}}
\newcommand{\benpo}[1]{{\color{violet}{(\textbf{Ben \& Po:} #1)}}}
\definecolor{darkBrickRed}{rgb}{.70,.13,.16}

\else
\newcommand{\fed}[1]{}
\newcommand{\benpo}[1]{}
\newcommand{\omer}[1]{}

\fi

\newcommand{\nobarfrac}{\genfrac{}{}{0pt}{}}

\newtheorem{theorem}{Theorem}
\newtheorem*{theorem*}{Theorem}
\newtheorem{lemma}{Lemma}
\newtheorem*{lemma*}{Lemma}

\newtheorem{corollary}{Corollary}

\newtheorem{proposition}{Proposition}

\newtheorem{axiom}{Axiom}

\theoremstyle{definition}

\theoremstyle{remark}

\newcommand{\bP}{\mathbb{P}}

\newcommand{\mP}{\mathbb{P}}

\newcommand{\R}{\mathbb{R}}

\newcommand{\one}{\mathds{1}}

\def\dd{\mathrm{d}}
\def\ee{\mathrm{e}}

\def\cA{\mathcal{A}}
\def\cB{\mathcal{B}}

\def\Z{\mathbb{Z}}
\def\N{\mathbb{N}}
\def\Q{\mathbb{Q}}

\def\probit{\mathrm{Probit}}
\def\sprobit{\mathrm{SProbit}}
\def\mnl{\mathrm{MNL}}
\def\lpcm{\mathrm{LPCM}}
\def\cO{\mathcal{O}}
\def\cM{\mathcal{M}}
\def\cT{\mathcal{T}}

\newcommand{\ml}{\mathrm{RCL}^\mu}
\newcommand{\rcl}{\textrm{RCL}}

\DeclarePairedDelimiter\ceil{\lceil}{\rceil}

\usepackage{xparse}
\usepackage{mleftright}

\DeclareDocumentCommand\Pr{ m g }{\ensuremath{
    {   \IfNoValueTF {#2}
      {\mathbb{P}\mleft[{#1}\mright]}
      {\mathbb{P}\mleft[{#1}\,\middle\vert\,{#2}\mright]}%
    }
}}
\DeclareDocumentCommand\E{ m g }{\ensuremath{
    {   \IfNoValueTF {#2}
      {\mathbb{E}\mleft[{#1}\mright]}
      {\mathbb{E}\mleft[{#1}\,\middle\vert\,{#2}\mright]}%
    }
}}

\title{\Large A Separability Foundation for Random Coefficients Logit\thanks{Previous versions of this paper were circulated under the titles ``Decomposable Stochastic Choice'' and ``Independence of Irrelevant Decisions in Stochastic Choice.'' We benefited from numerous suggestions and comments from our colleagues. We are grateful
to Marina Agranov, Victor Aguiar, Marco Bernardi, Alexander Bloedel, Peter Caradonna, Gabriel Carroll, Daniel Chen, Geoffroy de Clippel, Federico Echenique, Francesco Fabbri, Drew Fudenberg, Michael Gibilisco, Alexander Guterman, Wade Hann-Caruthers, Jakub Kastl, Michael Keane, Annie Liang, Daniel Litt, Jay Lu,
 Yusufcan Masatlioglu, Jeffrey Mensch, Stephen Morris, Pietro Ortoleva, Ariel Pakes, Tom Palfrey, Marcin P\c{e}ski, Luciano Pomatto, Gil Refael, Philipp Sadowski, Todd Sarver, Ilya Segal,
Barry Simon, Stanislav Smirnov, Tomasz Strzalecki, William Thomson, Aleh Tsyvinski, Christopher Turansick, Johan Ugander, Shoshana Vasserman, Ryan Webb, and Leeat Yariv.%, and seminar participants at
}
}

\author{Fedor Sandomirskiy\thanks{Princeton University. Email: fsandomi@princeton.edu} \and
        Po Hyun Sung\thanks{Caltech. Email: psung@caltech.edu} \and
        Omer Tamuz\thanks{Caltech. Email: tamuz@caltech.edu. Omer Tamuz was supported by a National Science Foundation CAREER award (DMS-1944153).} \and
        Ben Wincelberg\thanks{Caltech. Email: bwincelb@caltech.edu}}

\begin{document}
\maketitle
\begin{abstract}
We study stochastic choice across decision problems, each represented as a menu of action labels paired with observable outcome vectors. We propose a consistency condition for behavior in decision problems composed of two separable components: choice probabilities must agree with those obtained when each component is considered in isolation. Together with monotonicity and continuity, this separability requirement characterizes the family of random coefficients logit rules.

\end{abstract}

\section{Introduction}

Consider an analyst who studies individuals choosing among different actions, each yielding an observable outcome vector. 
From the analyst's perspective, choice behavior is stochastic: the population of decision makers may be heterogeneous, and, furthermore, they may choose actions with dominated outcomes for various reasons, including cognitive limitations, errors in the decision-making process, and random shocks unobserved by the analyst.
The analyst would like to develop a model to predict out-of-sample choice probabilities. In this paper we use an axiomatic approach to restrict the analyst's set of models. We show that random coefficients logit, a commonly used model, is unique in satisfying three simple axioms.\footnote{Random coefficients logit is a workhorse of empirical IO \citep{nevo2000practitioner,  moon2018estimation, gandhi2021empirical}, popularized by the seminal paper of \cite*{berry1995automobile}. It belongs to the broader class of mixed logit models, although some empirical work uses the two terms interchangeably \citep[see, e.g., the first paragraph in][]{nevo2000practitioner}. Following \cite{strzalecki2025stochastic}, we reserve the term mixed logit for mixtures over unrestricted utility functions and use random coefficients logit for mixtures of multinomial logit rules with linear utilities. See the literature review and \S\ref{sec:approx} for further discussion.}

We model a single decision instance as a \emph{menu} $(A,o)$ consisting of a finite set of actions $A$, each associated with an outcome $o(a) \in \R^n$. Each dimension of the outcome is a quantity that can be added (i.e., an extensive quantity) such as money, time, fuel, and so on. Following the discrete choice literature \citep{McFadden1981,strzalecki2025stochastic}, the actions are treated as mere labels, and the outcomes are assumed to carry all the information important to the decision maker. For example, consider the setting of a rideshare driver who faces the decision of which of several trips to accept. Each menu corresponds to a choice at a particular time. The actions in a menu are (meaningless) labels of trips, and the outcome associated with each action is a tuple $(r,-t,-g,-m)$ corresponding to the revenue from the trip, the time the trip will take, the gas it will consume and the mileage it will cover.  More broadly, the decision maker can be thought of as a producer, facing the choice of which good to produce. Alternatively, the decision maker is a consumer, and outcomes represent consumption bundles. In either case, outcomes can incorporate risk, for instance through the variance of revenue or other uncertain quantities.

A stochastic choice rule assigns to each menu a probability distribution over the set of actions, which we interpret as the predicted choice probabilities. The collection of stochastic choice rules is a rich, non-parametric family that gives rise to the problem of model selection. We restrict this family by considering three axioms: \emph{monotonicity}, \emph{continuity}, and \emph{independence of irrelevant decisions} (IID).

Our monotonicity axiom requires that an action that yields a dominating outcome (i.e., higher in every dimension) is taken with higher probability. This axiom places an ordinal restriction on the choice probabilities within a menu but places no restrictions across menus. Continuity is the technical assumption that a small change in outcomes leads to a small change in choice probabilities.

Our main axiom is independence of irrelevant decisions (IID). It is a condition imposed on additively separable menus, or, as we shall call them, \emph{product menus}. Such menus represent situations where multiple choices are made together, and the outcome from one choice does not affect the outcome from another. Suppose that an individual has to choose one of the two actions from $A_1 = \{a,b\}$, and also one of the two actions from $A_2 = \{s,t\}$. We can think of these two choices as a combined choice in a menu with action set $A = \{(a,s),(a,t),(b,s),(b,t)\}$. When the outcome $o(a_1,a_2)$ of each combined action is a sum $o_1(a_1)+o_2(a_2)$ of outcomes of its two components, we call $(A,o)$ a product menu. In the rideshare example, a product menu can correspond to the combination of two decisions at different points in time, where additive separability is natural since revenue, time, fuel, and mileage all aggregate across trips.
Of course, not all menus of combined choices are product menus, as there may be interactions between the choices that make them non-separable.

IID requires that choice probabilities over $A_1$ are the same, regardless of whether $(A_1,o_1)$ is faced in isolation or as part of the product menu $(A,o)$. The axiom does not rule out correlation across separable choices: a driver who favors shorter, less lucrative rides in the first trip may exhibit the same tendency in the
second.
Instead, IID captures a weak sense in which separability translates to behavior: decision makers' behavior in separable choices is consistent with their behavior when these choices are made in isolation. Importantly, for non-separable choices, where the menu is not a product menu, IID imposes no restrictions.

From the point of view of an analyst (say, employed by the rideshare company) who wants to predict the driver's choices, the IID assumption means that a stochastic choice rule will give the same prediction about a particular choice, regardless of whether it is considered together with other unrelated choices. Without IID, a model may provide different predictions for the driver's first choice of the day, depending on whether or not the analyst included later choices in the model. In a sense, when an analyst routinely excludes unrelated choices from their model, they are implicitly assuming that IID holds.

Our main result is that random coefficients logit rules are the only ones that satisfy monotonicity, continuity and IID (Theorem~\ref{thm:main}). Moreover, the coefficient distribution is identical across all menus. Thus, even though the IID axiom only restricts predictions for product menus, its conjunction with monotonicity and continuity implies that all choices---including in non-product menus---are made according to the same rule.

This result provides a simple foundation for this widely used choice rule. It also shows that one of the axioms is violated by any other  
stochastic choice rule, including 
one-shot probit and separable probit.
Indeed, 
one-shot probit violates IID, while separable probit violates monotonicity (see \S\ref{sec:separable}). 
The theorem thus highlights that  modeling even one decision instance with a rule that is inconsistent with  random coefficients logit carries hidden global assumptions. Regardless of how behavior is modeled on other menus, such a decision maker must either be influenced by the presence of irrelevant decisions (violating IID), fail to choose better actions with higher probability (violating monotonicity), or be extremely sensitive to outcomes (violating continuity).

In our main result we model separability by the addition of the outcomes in~$\R^n$. In~\S\ref{sec_general} we carry out a similar exercise, in a setting with an abstract outcome space and an abstract operation that captures separability. This highlights that it is separability---rather than the particular additive structure of~$\R^n$---that is the main driver of our results.

\subsection{Related Literature}

In this paper we provide an axiomatization of random coefficients logit. This term is often conflated with mixed logit, and we follow \cite*{strzalecki2025stochastic} to distinguish the two. Mixed logit allows unrestricted utility heterogeneity and can approximate any random utility model \citep{mcfadden2000mixed}, so axiomatizing mixed logit is close in spirit to axiomatizing stochastic rationality more generally \citep{manski1977structure, falmagne1978representation, mcfadden1990stochastic,clark1996random}. A characterization of mixed logit along these lines was obtained by \cite{saito2017axiomatizations}. 

Random coefficients logit restricts heterogeneity to linear utility functions over outcomes in $\R^n$. This model has become the workhorse of empirical research owing to its parsimony and flexibility: it allows for structured taste heterogeneity and combines this with tractability and identification
\citep[see, e.g.,][]{berry1995automobile,nevo2000practitioner,  Train_2003, fox2012random, moon2018estimation, gandhi2021empirical}.
To our knowledge, this paper is the first axiomatic characterization of this economically central subclass of mixed logit rules.

The restriction to linear utilities is substantive: linear-utility mixtures do not span all random utility behaviors unless the number of alternatives is small relative to the outcome dimension \citep{saito2017axiomatizations,lu2025mixed}. Such mixtures give rise to linear pure characteristic models (LPCM), which we discuss further in \S\ref{sec:approx} alongside a characterization of a related class by \cite{gul2006random}.  

In contrast to random coefficients logit, there is a large literature characterizing multinomial logit. Its early popularity stemmed from  analytical tractability and its microfoundation as a random-utility model with Gumbel-distributed shocks \citep*{luce1965utility, mcfadden1974conditional}. According to \cite*{luce1959individual}, a choice rule exhibits independence of irrelevant alternatives (IIA) if the relative probabilities for any subset of alternatives do not depend on the presence of other alternatives in the choice set. \cite*{luce1959individual} demonstrated that any behavior satisfying IIA can be generated by multinomial logit for some choice of (possibly nonlinear) utilities. Variants of IIA have since been used to characterize various generalizations of multinomial logit \citep*[see, e.g.,][]{gul2014random, echenique2018perception, ahn2018path, kovach2022behavioral}. Versions of multinomial logit with linear utilities can be characterized by augmenting IIA-type requirements with additional axioms pinning down the exponential dependence of likelihood ratios on outcomes \citep*{yellott1977relationship, cerreia2021axiomatic, breitmoser2021axiomatic, cerreiamultinomial}. Unlike IIA, which constrains behavior across nearly identical menus such as those generated by duplicating an action as in \citep*{debreu1960review}, IID places no restrictions in such settings and applies only to decisions made in separate, non-overlapping contexts.

A separate literature microfounds multinomial logit and related rules through information or cognitive frictions. \cite*{matvejka2015rational} show that multinomial logit captures the behavior of a utility-maximizing individual with entropy-based attention costs \citep*{sims2003implications}. \cite*{woodford2014stochastic} and \cite*{mattsson2002probabilistic} derive related results for binary choices and costly effort, respectively. \cite*{steiner2017rational} obtain an entropy-cost characterization of dynamic logit; see also \cite*{fudenberg2015dynamic}. In a multiperiod context, \cite*{fudenberg2024limited} show that mixed probit describes the limiting choice of an agent with bounded memory and Gaussian information.  In contrast to this literature, our paper is agnostic about the origin of logit behavior, and instead shows that it is the only form of behavior compatible with consistent choices across separable decision problems.

Our analysis of separability in general outcome spaces (\S\ref{sec_general}) connects to a classical literature showing that choice independence across dimensions yields additively separable utility \citep*{debreu1959topological, fishburn1965independence, segal1992additively, wakker2013additive,cho2022add}.
Similarly, we demonstrate that choice independence across product menus implies that choices are driven by a utility function that is additively separable across the product components.

A closely related line of  work on separability focuses on consistency of behavior across agents or periods. \cite*{chambers2021correlated} and \cite*{kashaev2024entangled} consider the choice behavior of two agents (or of a single agent over two periods) and study its separability, asking whether a joint distribution over choices can be rationalized by a single distribution over utility pairs. Similar cross-period separability restrictions appear in dynamic random-utility frameworks \citep*{frick2019dynamic, li2021dynamic, kashaev2023dynamic}. These papers share two features with ours: they impose consistency across joint decisions, and they conclude that separability pins down a stable underlying stochastic preference governing such choices.
The key conceptual difference is that  our separability requirement sharply disciplines  model selection, leaving only random coefficients logit.  A version of our separability-based axiomatic approach is explored by \cite*{sandomirskiy2025monotonicity} in a multi-agent strategic context, where it is used to axiomatize Nash and quantal response equilibria along with new solution concepts.

\section{Model}\label{sec_model}
Let $\cA$ be a universal set of actions.
We assume that this set is non-empty and closed under the operation of forming ordered pairs. In other words, if $a_1,a_2 \in \cA$ then the pair $(a_1,a_2)$ is also an element of $\cA$. We interpret $(a_1,a_2)$ as a new action label representing the compound action of taking $a_1$ and $a_2$ together.
Note that this condition implies that $\cA$ is infinite.\footnote{Indeed, if $a \in \cA$, then $(a,a) \in \cA$, and hence $((a,a),a) \in \cA$, and so on.} We further assume that $\cA$ is countable.

 The set of possible outcomes of a decision is denoted by~$\cO$. A single decision instance is represented by a \emph{menu} $(A,o)$, where $A \subset \cA$ is a non-empty finite set of possible actions and~$o\colon A\to \cO$ assigns an outcome to each action. 
Note that the outcome of an action encapsulates all the information about this action relevant to the decision-maker. In contrast, the name of the action is just a label. In particular, the same action may appear in different menus and result in different outcomes.

The key setting we study is $\cO = \R^n$; in \S\ref{sec_general} we generalize beyond $\R^n$.  To build intuition it is useful to keep in mind the example of $\cO=\R$. This benchmark outcome space can be used to model decision-makers who compare actions by a single number, such as their monetary reward---and we accordingly refer to outcomes as payoffs. We use $\cO=\R$ in the examples used to illustrate the definitions of this section.

We display a menu by showing each action's outcome below it. For example,
\begin{align*}
    (A,o) &= \left\{\nobarfrac{\text{a}}{3.14}~~~\nobarfrac{\text{b}}{-17}\right\}
\end{align*}
is a menu with two actions, choosing $a$ or $b$, with the former having an outcome  of $3.14$ and the latter having an outcome of $-17$.

The collection of all menus is denoted by $\cM$. Thus, when $\cO=\R^n$,
$$  \cM=\big\{(A,o) \colon A \subset \cA \ \ \text{ is finite and non-empty,} \  o\colon A \to \R^n\big\}.
$$
Since actions and outcomes are specified separately, two menus may have the same action set but assign different outcomes to its elements.

This framework sits between two standard stochastic-choice domains. In the classical random-utility tradition, a decision problem is a set of primitive alternatives, and all choice-relevant information is carried by the identities of those alternatives \citep[see, e.g.,][]{block1959random,luce1965utility,
falmagne1978representation}. In pure-characteristics models, alternatives are represented directly as points in $\R^n$, so alternatives with the same vector are identified at the level of the model \citep[see, e.g.,][]{saito2017axiomatizations,
lu2025mixed}. Our framework keeps these two roles distinct. As in
characteristics-based models, the relevant information about action $a$ in menu $(A,o)$ is contained in $o(a)$. However, we do not reduce the menu to the set of outcomes $o(A) \subset \R^n$. Retaining action labels lets us form compound actions, such as $(a_1,a_2)$, and relate actions across different menus. The roles played by actions and outcomes in our framework fit the game-theoretic tradition in which the actions available to a player are mere labels and only the
payoffs matter; indeed, for $\cO=\R$, a menu $(A,o)$ can be viewed as a one-player normal-form game with action set $A$ and payoff function $o$.

\smallskip
A \emph{stochastic choice rule} is a map $\Phi$ that assigns to each menu $(A,o)\in \cM$ a probability distribution over $A$. We denote by $\Phi(a \mid A,o)$ the probability that $\Phi(A,o)$ assigns to $a \in A$. We think of $\Phi$ as describing or predicting the choices of a  decision maker across different situations.

One family of widely used stochastic choice rules consists of the \emph{independent additive random utility}  models (IARU), which are given by
\begin{align*}
    \mathrm{IARU}^u(a \mid A,o) = \bP\Big[u(o(a))+\varepsilon_a=\max_{b\in A} u(o(b))+\varepsilon_b\Big],
\end{align*}
where $u \colon \R^n \to \R$ is a utility function and $(\varepsilon_b)_{b\in A}$ are independent shocks with a common continuous CDF $F$.

When these shocks follow the Gumbel distribution, i.e., $F(t)=\exp(-\exp(-t))$, this is the \emph{multinomial logit rule} (MNL), which is given by
\begin{align*}
    \mnl^u(a \mid A,o) = \frac{\exp({u(o(a))})}{\sum_{b \in A}\exp({u(o(b))})}.
\end{align*}
A particularly important subclass is \emph{linear multinomial logit}, 
in which $u(x) = \beta \cdot x$ for some $\beta \in \R^n$:
\begin{align*}
    \mnl^\beta(a \mid A,o) = \frac{\exp({\beta \cdot o(a)})}{\sum_{b \in A}\exp({\beta \cdot o(b)})}.
\end{align*}
Multinomial logit is commonly used in the empirical literature for its computational tractability, but it does not allow for random taste variation and various substitution patterns \citep*{Train_2003}. These limitations are overcome by the \emph{random coefficients logit rule} (RCL), which is a weighted average of linear multinomial logit rules, given by \begin{align*}
    \ml(a \mid A,o) = \int   \mnl^\beta(a \mid A,o) \,\dd \mu(\beta).
\end{align*}
It is parameterized by a probability measure $\mu$ over the logit parameter $\beta\in \R^n$.  More generally than random coefficients logit, one can consider mixtures of (non-linear) multinomial logit rules, which is the class of mixed logit rules. In the empirical literature these terms are often conflated \citep{nevo2000practitioner,gandhi2021empirical}, but in our setting the difference is important, as we discuss in \S\ref{sec:approx}.

\medskip

We consider several properties of stochastic choice rules. The first one is monotonicity. We consider the set of outcomes $\R^n$ together with the usual component-wise partial order $\geq$, so that higher outcomes are universally preferred.\footnote{Our results hold more generally for any partial order, including the trivial one in which each outcome is only comparable to itself. In \S\ref{sec_general} we formally consider general partial orders.}  Monotonicity reflects this preference. In particular, it limits the possible influence of action labels on choice behavior.\footnote{While our model is motivated by the view that only outcomes matter to the decision maker, we do not directly impose this as an axiom or hardwire it into the definition of a choice rule. Nevertheless, dependence on labels will be ruled out by the combination of the other axioms we impose.
}

\begin{axiom}[Monotonicity]
\label{ax:monotone}
A rule $\Phi$ is \textbf{monotone} if for any menu $(A,o)$ in $\cM$ and any $a,a' \in A$ such that $o(a) \geq o(a')$ it holds that $\Phi(a \mid A,o) \geq \Phi(a' \mid A,o)$.
\end{axiom}
  Note that this axiom does not impose any constraints across menus, but only within a given menu.

  Monotonicity is satisfied by all IARU models with monotone utilities. Moreover, the class of stochastic choice rules that satisfy monotonicity is convex, i.e., a mixture of monotone rules is monotone. In particular, random coefficients logit rules whose coefficient distributions are supported on $\R_{\geq 0}^n$ satisfy monotonicity.

For a fixed set of actions~$A$, we say that a sequence of menus $(A,o_n)$ converges to~$(A,o)$ if $\lim_n o_n(a)=o(a)$ for all $a \in A$. 
\begin{axiom}[Continuity]
  \label{ax:cont}
 A rule $\Phi$ is \textbf{continuous} if for any sequence of menus  $(A,o_n)$ from $\cM$ converging to $(A,o)$, we have $\lim_n\Phi(a \mid A,o_n)=\Phi(a \mid A,o)$ for all $a \in A$.
\end{axiom}
Alternatively put, continuity stipulates that very small changes in the outcomes result in very small changes in choice probabilities. This axiom excludes stochastic choice rules that describe individuals who pay excessive attention to even negligible differences in outcomes. For instance, it is violated by rules for $\cO=\R$ that always select one of the highest-outcome actions, regardless of how small the advantage is. Nevertheless, continuity is a common modeling choice made for good reason: people do not always choose the dominant action, especially when the difference between outcomes is minuscule. 

\medskip

Our main axioms concern choice rule predictions on menus that represent multiple decisions with additively separable outcomes. 
We say that $(A,o)$ is a \emph{product menu} if 
\begin{equation}\label{eq_addtive_outcomes}
A=A_1 \times A_2\quad \text{and}\quad o(a_1,a_2) = o_1(a_1)+o_2(a_2).
\end{equation}
That is, $A$ consists of action pairs $a=(a_1, a_2)$ with $a_1 \in A_1$ and $a_2 \in A_2$, and the outcome assigned to each pair is additively separable. We write $(A,o) = (A_1,o_1) \otimes (A_2,o_2)$ and refer to $(A,o)$ as the product of $(A_1,o_1)$ and $(A_2,o_2)$.

For example, suppose an experimenter runs two tests consecutively on the same subject. In the first test, the subject chooses between two actions $a$ and $b$, and receives payoff $0$ or $1$, respectively. In the second, the subject chooses between $s$ and~$t$, and again receives $0$ or $1$. Indeed, many experiments contain comprehension questions and pay subjects for each correct answer.

The first test would be well-modeled by the menu $(A_1,o_1)$, with $A_1=\{a,b\}$ and $o_1(a)=0$, $o_1(b)=1$. Likewise, the second test would be well-modeled by $(A_2,o_2)$, with $A_2=\{s,t\}$ and $o_2(s)=0$, $o_2(t)=1$. The joint decision the subject faces is the product menu
\begin{align*}
    (A,o) = \left\{\nobarfrac{\text{a}}{0}~~~\nobarfrac{\text{b}}{1}\right\}  \otimes \left\{\nobarfrac{\text{s}}{0}~~~\nobarfrac{\text{t}}{1}\right\} 
 = \left\{\nobarfrac{\text{(a,s)}}{0}~~~\nobarfrac{\text{(a,t)}}{1}~~~\nobarfrac{\text{(b,s)}}{1}~~~\nobarfrac{\text{(b,t)}}{2}\right\} 
 =\left\{\begin{array}{c|cc}
   & s & t \\
\hline
a & 0 & 1 \\
b & 1 & 2 \\
\end{array}\right\}.
\end{align*}

Alternatively, the experimenter could only award the subject if both questions are answered correctly. This experiment would be well-modeled by the menu 
\begin{align*}
    (A,o') =  \left\{\nobarfrac{\text{(a,s)}}{0}~~~\nobarfrac{\text{(a,t)}}{0}~~~\nobarfrac{\text{(b,s)}}{0}~~~\nobarfrac{\text{(b,t)}}{1}\right\} = \left\{
\begin{array}{c|cc}
   & s & t \\
\hline
a & 0 & 0 \\
b & 0 & 1 \\
\end{array}
\right\}.
\end{align*}
This is not a product menu, even though the action set is a product set (indeed, the same product set), because the outcomes are not additively separable.  This example highlights that we do not assume that arbitrary combinations of decisions give rise to product menus. While any such combination involves taking the Cartesian product of the underlying action sets, the associated outcomes need not take the separable form~\eqref{eq_addtive_outcomes}.

For an example with $\cO = \R^n$, consider the setting of a rideshare driver who faces decisions of which of several trips to accept. Each menu corresponds to a choice at a particular time. The actions in a menu are (meaningless) labels of trips, and the outcome associated to each action is a tuple $(r,-t,-g,-m)$ corresponding to the revenue from the trip, the time the trip will take, the gas it will consume and the mileage it will cover.  A product menu corresponds to the combined decision in two subsequent rides, where separability is natural, as all the above quantities are added across trips. Beyond the specific example of a driver, the decision maker can be thought of as a business owner, facing the choice of which jobs to accept, with a vector of various quantities describing the possible outcomes. Quantities could also include risk (e.g., the variance of the revenue or of other uncertain outcomes). A similar interpretation  is that the decision maker is a consumer, and outcomes represent consumption bundles.

A multidimensional outcome space can also capture choice under ambiguity. Let $\Theta$ be a finite set of states. An outcome $x$ is a function $x\colon \Theta\to \R$, specifying state-contingent payoffs. That is, $x$ is a Savage act and the space of all outcomes can be identified with $\cO = \R^\Theta$. A decision-maker is ambiguous about the state $\theta\in \Theta$ and so may take into account all the possible values $x_\theta$. Product menus then correspond to combinations of two decisions leading to separable state-contingent payoffs.

\medskip

Our main axiom, IID, restricts the choice rule only on product menus. It requires that choice predictions for a product menu be consistent with predictions for each component considered separately. We begin by discussing a stronger assumption, which additionally requires statistical independence of choices across the two dimensions of a product menu.
\begin{axiom}[Decomposability]
  \label{ax:frame}
 A rule $\Phi$ satisfies \textbf{decomposability} if for all 
product menus $(A,o)=(A_1,o_1)\otimes(A_2,o_2)$ in $\cM$, it holds that
 \begin{align*}
     \Phi((a_1,a_2)\mid A,o)= \Phi(a_1 \mid A_1,o_1) \cdot \Phi(a_2 \mid A_2,o_2)
 \end{align*}
 for all $(a_1,a_2)\in A$.
\end{axiom}

For example, suppose we observe the choice probabilities of $(A_1,o_1)$ and $(A_2,o_2)$ to be
    \begin{align*}
      \Phi\left(\nobarfrac{\text{a}}{0}~~~\nobarfrac{\text{b}}{1}\right)=\Phi\left(\nobarfrac{\text{s}}{0}~~~\nobarfrac{\text{t}}{1}\right)
      = (1/3,2/3).
    \end{align*}

Decomposability requires that for the product menu,
     \begin{align*}
 \Phi\left(\begin{array}{c|cc}
   & s & t \\
\hline
a & 0 & 1 \\
b & 1 & 2 \\
\end{array}\right)
      = \begin{array}{c|cc}
   & s & t \\
\hline
a & 1/9 & 2/9 \\
b & 2/9 & 4/9 \\
\end{array}.
    \end{align*}

    Decomposability means that, for product menus, the prediction is the same as when that decision is made in isolation. Moreover, the predicted distribution is statistically independent across the two dimensions. In the experimental lab example, this would imply that subjects choose the wrong answer independently in the two questions they are asked, when they are rewarded separately for each correct answer. Decomposability imposes no restriction on predicted behavior in situations where subjects are rewarded only for answering both questions correctly, as such situations do not correspond to product menus.  
    
    While decomposability is a simple separability assumption, in some settings its independence component may be unrealistic, especially in the presence of unobservable heterogeneity.    In the lab example, one could expect that subjects who answer the first question correctly are more likely to also answer the second correctly. In the driver example, it is natural to expect that the choices a driver makes through the day are not necessarily independent of each other.

The IID axiom is a weakening of decomposability, allowing for such correlations while maintaining consistency with predictions for the components. Given a product menu $(A,o)=(A_1,o_1) \otimes (A_2,o_2)$, we denote the marginal choice probability of $a_1\in A_1$ by
\begin{align*}
    \Phi(a_1 \mid A,o) = \sum_{a_2 \in A_2} \Phi\left((a_1,a_2) \mid A,o\right).
\end{align*}
\begin{axiom}[IID]
  \label{ax:IID}
 A rule $\Phi$ satisfies \textbf{independence of irrelevant decisions} if for all product menus $(A,o)=(A_1,o_1)\otimes(A_2,o_2)$ in $\cM$, it holds that
 \begin{align}\label{eq_iid}
\Phi(a_1 \mid A,o) = \Phi(a_1 \mid A_1,o_1)
 \end{align}
 for all $a_1 \in A_1$.
\end{axiom}

For example, for the menu $(A,o) = (A_1,o_1) \otimes (A_2,o_2)$ above, IID implies that if \begin{align*}
      \Phi\left(\nobarfrac{\text{a}}{0}~~~\nobarfrac{\text{b}}{1}\right)=\Phi\left(\nobarfrac{\text{s}}{0}~~~\nobarfrac{\text{t}}{1}\right)
      = (1/3,2/3),
    \end{align*} then
\begin{align*}
    \Phi(a \mid A,o) = \Phi((a,s) \mid A,o) + \Phi((a,t) \mid A,o)=\frac{1}{3}.
\end{align*} Thus, IID allows for predictions such as
  \begin{align*}
 \Phi\left(\begin{array}{c|cc}
   & s & t \\
\hline
a & 0 & 1 \\
b & 1 & 2 \\
\end{array}\right)
      = \begin{array}{c|cc}
   & s & t \\
\hline
a & 1/6 & 1/6 \\
b & 1/6 & 3/6 \\
\end{array},
    \end{align*} 
which is not a product measure, but it does not allow
\begin{align*}
 \Phi\left(\begin{array}{c|cc}
   & s & t \\
\hline
a & 0 & 1 \\
b & 1 & 2 \\
\end{array}\right)
      = \begin{array}{c|cc}
   & s & t \\
\hline
a & 1/8 & 2/8 \\
b & 2/8 & 3/8 \\
\end{array}.
    \end{align*}
Like decomposability, the IID axiom imposes no restrictions on choice rule predictions for non-product menus such as $(A,o')$.

IID means that  
predictions are independent of the inclusion of an unrelated menu into the analysis. In other words, for product menus, the predicted choice frequency of an action in the first dimension is consistent with predictions when only that dimension is considered. Without the assumption of IID, a modeler would have to include all irrelevant decisions that the population faces in order to make an accurate prediction.

IID is satisfied by random coefficients logit, and in particular is satisfied by linear multinomial logit, which moreover satisfies decomposability. The class of rules that satisfy IID is very large. For instance, we can construct such a rule $\Phi$ by specifying any  choice distribution on each non-product menu, and then defining $\Phi$ on product menus to be the product measure given by $\Phi$ on each component menu. This construction will in fact satisfy decomposability, and is very flexible. 

A natural example of this sort is a class of models that we call \emph{separable IARU} ($\mathrm{SIARU}$), which is defined as follows. Fix a utility $u \colon \R^n \to \R$. For non-product menus,\footnote{For present purposes, these include products in which one of the menus is trivial, in the sense that it includes only one action.} it coincides with the one-shot rule $\mathrm{IARU}^u$. For product menus, it is given by
\begin{align}
    \label{eq:siaru}
    \mathrm{SIARU}^u((a_1,a_2)\mid A,o) 
    &= \mathrm{SIARU}^u(a_1\mid A_1,o_1)\cdot \mathrm{SIARU}^u(a_2 \mid A_2,o_2).
\end{align}
That is, choices are made independently in each dimension. For example, if a consumer makes two unrelated choices---e.g., purchasing a milk and a soap---they choose the best milk subject to some noise and the best soap subject to some additional noise, with the noises independent for the two choices. Such a model could also be applied in a dynamic setting: a consumer chooses milk both today and tomorrow, and faces independent shocks for the two periods.\footnote{We thank an anonymous referee for suggesting this example.} See~\S\ref{sec:separable} for a formal definition and discussion.

\subsection{Implications of the Axioms}\label{sec_axiom_implications}

The main result of this paper is a characterization of all monotone, continuous rules that satisfy IID. Before presenting our main result, we illustrate that while each of these axioms on its own is rather weak, together they have surprisingly strong implications, and already for $\cO = \R$.

For example, consider an analyst who is interested in choice probabilities in the menu
\begin{align*}
    (A,o) &= \left\{\nobarfrac{a_1}{0}~~~\nobarfrac{a_2}{2}~~~\nobarfrac{a_3}{7}\right\}.
\end{align*}
Suppose that the analyst observes the choice probability of the action $a_1$ to be $5\%$, i.e., $\Phi(a_1 \mid A,o) = 5\%$. 

Clearly, monotonicity implies $\Phi(a_3 \mid A,o) \geq 47.5\%$, since $\Phi(a_2 \mid A,o) + \Phi(a_3 \mid A,o) = 95\%$ and $\Phi(a_2 \mid A,o) \leq \Phi(a_3 \mid A,o)$. Without monotonicity, IID yields no constraints for $\Phi(a_3 \mid A,o)$, since $(A,o)$ is not a product menu (indeed, even decomposability has no further implications). Naively, the combination of monotonicity and IID does not seem to imply any further constraints. 

Surprisingly, this intuition is wrong. It turns out that if we assume that $\Phi$ is monotone and satisfies IID, then $\Phi(a_1 \mid A,o) = 5\%$ implies that $\Phi(a_3 \mid A,o) \geq  83.8\%$. This bound is tight: if we make the stronger assumption that $\Phi$ is monotone and decomposable, then $\Phi(a_1 \mid A,o) = 5\%$ implies that $\Phi(a_3 \mid A,o) =  83.8\%$. As we explain in detail below, this is a consequence of our main results. Furthermore, there is a unique $\Phi$ that is monotone, decomposable, continuous, and satisfies $\Phi(a_1 \mid A,o) = 5\%$; see Appendix~\S\ref{sec:restriction} for details.

\section{IID and Random Coefficients Logit}\label{sec_IID_mixed}

Consider the outcome space $\cO = \R^n$. For $\mu$ supported on $\R_{\geq 0}^n$, the random coefficients logit rule given by
\begin{align}\label{eq_ML}
    \ml = \int   \mnl^\beta \,\dd \mu(\beta)
\end{align}
 satisfies monotonicity, continuity, and IID. Our main theorem states that no other rules satisfy these axioms.
\begin{theorem}
\label{thm:main}
Let $\Phi$ satisfy monotonicity, continuity, and IID. Then $\Phi$ coincides with a random coefficients logit rule $\ml$ for some~$\mu$ supported on $\R_{\geq 0}^n$.
\end{theorem}

The theorem is proved in \S\ref{sec:thm_main} for $\cO = \R$ and the multidimensional case is covered in Appendix~\ref{app:th_multi}. Some of the core ideas are explained in \S\ref{sec:correlations}.

An interpretation of the theorem is that any decision maker satisfying the assumptions behaves as a stochastic utility maximizer with linear utility, random taste, and random noise level: the taste is encoded by the direction of $\beta$, while the noise level is inversely proportional to its magnitude. The variation in $\beta$ may reflect heterogeneity  across a population or variability of the internal state  of a single individual.  Conditional on each realization of $\beta$, choices follow a linear multinomial logit rule. The realized $\beta$ is unobservable and thus the observable choice takes the form of a mixture over such logits, with mixing distribution~$\mu$.  Notably, the fact that choice probabilities follow a random utility model emerges from the axioms and is not assumed a priori. 

In the one-dimensional case $\cO=\R$ where outcomes are interpreted as payoffs, $\beta \in \R_{\geq 0}$ determines the (random) noise level of a decision maker who is optimizing, but may stochastically choose a dominated outcome. In the example of the rideshare driver with outcomes $(r,-t,-g,-m)$ where the revenue $r$ is measured in dollars, we can write $\beta = \gamma \cdot p$, where $\gamma$ is in $\R_{\geq 0}$ and $p = (p_1,p_2,p_3,p_4)$ is a non-negative vector normalized to $p_1=1$. We can now interpret the components $p_2,p_3$ and $p_4$ as the prices that the driver assigns to their time, gas and mileage. The driver is stochastically optimizing their expected profit, with noise $\gamma$. In the choice under ambiguity example, where outcomes are vectors of payoffs indexed by state $\theta$, one can instead normalize $p$ to sum to one, recasting it as a probability measure over the state space $\Theta$. Under this normalization, the decision maker becomes a Bayesian stochastic expected payoff maximizer.

When some coordinates of the outcome vector represent costs rather than payoffs, it is natural to adjust the notion of monotonicity so that the decision maker favors actions with lower values in those coordinates.\footnote{For instance, if outcomes in the rideshare example are
written as $(r,t,g,m)$ rather than $(r,-t,-g,-m)$, it is natural to impose monotonicity with respect to the order $\geq'$ on $\R^4$ defined by $(r,t,g,m)\geq'(r',t',g',m')$ whenever $r\geq r'$,
$t\leq t'$, $g\leq g'$, and $m\leq m'$.} Theorem~\ref{thm:main} extends straightforwardly to arbitrary partial orders on $\R^n$, since its proof invokes the monotonicity axiom only to ensure that two actions with identical outcomes are chosen with equal probability. Accordingly, if monotonicity is imposed with respect to a given partial order~$\geq'$ on $\R^n$, the corresponding version of Theorem~\ref{thm:main} states that the rule coincides with $\rcl^\mu$ for a measure $\mu$ supported on the set of $\beta\in\R^n$ for which the linear function $x\to\beta\cdot x$ is $\geq'$-monotone. At the extreme, taking $\geq'$ to be the trivial order ($x\geq' y$ if and only if $x=y$) reduces monotonicity to the requirement that equal-outcome actions are
selected with equal probability, and $\mu$ can be supported on all of $\R^n$. We discuss
monotonicity with respect to general partial orders further in \S\ref{sec_general}.

The theorem highlights several additional consequences of the axioms, beyond explicitly characterizing the form of choice rules consistent with them.  First, while our definition of a stochastic choice rule and each axiom in isolation permit dependence on action labels, the axioms jointly rule this out: only the profile of outcomes in the menu can affect choice probabilities, eliminating framing effects. Second, 
although IID---the only axiom linking behavior across menus---places no restrictions on non-product menus by itself, its interaction with the other axioms pins down behavior on all menus. As a result, the same random coefficients logit rule governs choice even in compound decisions that are not representable as product menus---for example, those involving interrelated choices or shared constraints. 

\subsection{Mixed Logit, Random Coefficients Logit, and Universal Approximation}\label{sec:approx}

It is instructive to discuss the connection between our characterization of random coefficients logit and the universal approximation property of mixed logit rules established by \cite{mcfadden2000mixed}. Effectively, they show that any aggregate choice behavior of a population of rational individuals can be approximated with general mixed logit rules, i.e., mixtures of $\mnl^u$ over different utilities $u$, as long as one can choose utilities from a rich enough class such as all polynomials of high degree. 

This property of mixed logit highlights that it is an extremely flexible model, making it challenging to identify and estimate. In contrast, random coefficients logit---in which the utilities $u$ are restricted to be linear---is widely used because it is parsimonious and hence identifiable, while retaining sufficient flexibility to capture many phenomena.\footnote{Of course, if one expands the vector of outcomes associated with an action by adding extra coordinates equal to non-linear functions of the original outcome vector \citep[e.g., powers of other coordinates, as in][]{mcfadden2000mixed}, then the linear functions effectively describe previously non-linear ones. However, this hurts identification, and hence cannot be over-used \citep*[see, e.g.,][Assumption 2.1]{fox2012random}. Relatedly, the distinction between linear and non-linear utilities vanishes when the number of alternatives is smaller than the number of outcomes, as in \cite{saito2017axiomatizations}.}

Theorem~\ref{thm:main} does not characterize the entire class of mixed logit rules, but rather just the subclass of mixtures corresponding to linear utilities of the form $u(x)=\beta\cdot x$, which gives rise to the random coefficients logit rules. In particular, not all mixed logit rules satisfy IID: among the continuous and monotone ones, IID is satisfied only by random coefficients logit. 

An approximation result for random coefficients logit was explored by \cite{lu2025mixed}, who show that such mixtures can approximate any \emph{linear pure characteristic model} (LPCM). These rules represent the aggregate choice behavior of a population of rational individuals with linear utilities and are defined as follows
$$\lpcm^\mu(a\mid A,o)=\int \frac{\one_{a\in\arg\max_{b\in A} \beta\cdot o(b)}}{|\arg\max_{b\in A} \beta\cdot o(b)|}\dd\mu(\beta).$$
Since $\mnl^{t\beta}$ with $t\to +\infty$ approximates the choice of a rational individual with $u(x)=\beta\cdot x$, one can show that $\lpcm^\mu(a\mid A,o)$ can be approximated by mixing over such linear logits with $\beta$ distributed according to  $\mu$.
As limits of rules satisfying IID and monotonicity, $\lpcm^\mu$ with $\mu$ supported on $\R_{\geq 0}^n$ also satisfies these axioms (this can also be verified directly). However, $\lpcm^\mu$ violates the continuity assumption and thus is ruled out in Theorem~\ref{thm:main}.\footnote{A characterization of a class of rules similar to LPCM, though on a different domain, is obtained by \cite{gul2006random}. They consider a decision maker facing finite choice sets of lotteries and characterize random expected utility. Interpreting lottery weights as outcome vectors, random expected utility models become mathematically equivalent to LPCM. The key axioms are linearity, playing the role of the random-choice analogue of the von Neumann--Morgenstern independence axiom (equivalently, invariance to affine transformations of the outcome vectors); monotonicity across nested choice sets; and extremeness requiring that only choices maximizing some linear utility function receive positive weight. Random coefficients logit violates linearity because rescaling outcomes alters the relative magnitude of logit noise, and it violates extremeness because logit assigns positive probability to every choice, including dominated ones.}

\section{Correlations, Decomposability, and Linear Multinomial Logit}
\label{sec:correlations}
The IID axiom does not preclude correlations between separable choices. In this section we tackle two questions. First, we want to better understand what types of correlations IID allows, once it is taken in conjunction with continuity and monotonicity. Second, we would like to understand the implications of requiring that there are no correlations, i.e., replacing IID with the stronger axiom of decomposability.

Our first observation is that in the one-dimensional $\cO =\R$ case, rules satisfying IID, monotonicity and continuity (i.e., random coefficients logit rules, by Theorem~\ref{thm:main}) cannot display negative correlations. Since variability in choice in the one-dimensional case stems from mistakes rather than taste variations, we interpret this result as stating that mistakes must be non-negatively correlated. 

Formally, we define correlation as follows. Given a menu $(A,o)$ and a function $f \colon A \to \R$ we will write $\Phi(f(a) \mid A,o) = \sum_a f(a)\Phi(a\mid A,o)$ for the expectation of $f$ under the probability measure $\Phi(\,\cdot\,|A,o)$. A stochastic choice rule $\Phi$ exhibits 
     \emph{non-negative correlation}    
    on a product menu $(A,o)=(A_1,o_1)\otimes (A_2,o_2)$ if  \begin{equation}\label{eq:pos_cor}
    \Phi\Big(o_1(a_1)\cdot o_2(a_2)\,\Big|\, A,o\Big) \geq \Phi\Big(o_1(a_1)\,\Big|\,  A,o\Big)\cdot\Phi\Big(o_2(a_2)\,\Big|\,  A,o\Big).
    \end{equation}
We say that $\Phi$ exhibits \emph{zero correlation} if \eqref{eq:pos_cor} holds with equality. A menu $(A,o)$ is said to be non-trivial if $o$ is non-constant. 

In the example of the lab experiment in which every correct answer yields a payoff of one, non-negative correlation means that when a subject answers the first question correctly they are (weakly) more likely to answer the second one correctly.

\begin{proposition}\label{prop:pos_cor}
For any distribution $\mu$ on $\R$, $\ml$ exhibits non-negative correlation on every product menu.  Moreover, $\ml$ exhibits zero correlation on a product of non-trivial menus if and only if $\mu$ is a Dirac measure. 
\end{proposition}
Intuitively, because the mixing distribution $\mu$ is fixed across all menus, lower noise levels (and thus better decisions) tend to occur simultaneously across the components of a product menu. This pattern is a joint consequence  of the three axioms rather than of IID, which allows for arbitrary correlation in product menus. 

The proof of Proposition~\ref{prop:pos_cor} appears in \S\ref{ap_prop1}, and is straightforward. Together with Theorem~\ref{thm:main}, we conclude that every rule that satisfies our axioms has non-negative correlation, which is not immediately evident from the axioms. In~\S\ref{sec:separable}, we provide an example demonstrating that probit models do not always exhibit non-negative correlation, highlighting the connection between this property and our axioms. 

\bigskip

Next, we ask which of the rules satisfying our axioms furthermore satisfy decomposability, i.e., exhibit no correlations across product menus. The following result holds for outcomes $\cO=\R^n$.
\begin{proposition}\label{prop:mnl}
    Let $\Phi$ satisfy monotonicity, continuity and decomposability. Then~$\Phi$ coincides with a linear multinomial logit rule $\mnl^\beta$ for some~$\beta \in \R_{\geq 0}^n$.
\end{proposition}
 These propositions imply the well-known fact \citep*[see][]{fox2012random} that a non-degenerate mixture of multinomial logit rules cannot itself be a multinomial logit rule. This yields a geometric interpretation of Theorem~\ref{thm:main}: the set of choice rules satisfying monotonicity, continuity, and IID is convex, with multinomial logit rules as its extreme points.\footnote{An extreme point of a convex set is one that cannot be written as a non-trivial convex combination of others. In this light, representation~\eqref{eq_ML} mirrors the Choquet theorem: any choice rule satisfying the axioms can be expressed as a mixture of extreme points. \cite*{fox2012random} show that this representation is unique, that is $\rcl^\mu = \rcl^{\mu'}$ implies $\mu = \mu'$, so the space of such rules forms an infinite-dimensional simplex. This fact also follows from Proposition~\ref{prop_identification} in \S\ref{sec:thm_main}, which is a part of the proof of our main theorem. In fact, this proposition shows that $\mu$ is already identified by the choice probabilities on menus with integer outcomes.} 

\bigskip

The proof of Proposition~\ref{prop:mnl} appears in Appendix~\ref{app_mnl_proof}, and uses Theorem~\ref{thm:main} and Proposition~\ref{prop:pos_cor}. We now sketch a direct proof for Proposition~\ref{prop:mnl} for the case $\cO=\R$. This also provides a simple explanation of the core ideas of the proof of our main result, Theorem~\ref{thm:main}.  The formal proof of Theorem~\ref{thm:main} follows a similar path and makes use of the De Finetti Theorem to show that $\Phi$ is a convex combination of decomposable rules.

Suppose $\Phi$ is decomposable and monotone. We show that $\Phi$ (restricted to menus with rational outcomes) is identified by its predictions for a single menu, and that this implies that $\Phi$ is a multinomial logit rule. Assume we know $\Phi(B,r)$ for 
\begin{align*}
  (B,r) &=
    \left\{\nobarfrac{b_0}{0}~~~\nobarfrac{b_1}{1}\right\}.
\end{align*}
For the sake of this proof sketch, suppose also that both $b_0$ and $b_1$ are chosen with positive probability. Our goal is to show how this knowledge pins down $\Phi$ on any given menu $(A,o)$. Since multinomial logit rules are monotone and decomposable, this will immediately imply that $\Phi$ is a multinomial logit rule, with the parameter $\beta$ chosen to 
agree with $\Phi$ on $(B,r)$.

For clarity, consider the particular example
\begin{equation*}
(A,o) =
    \left\{\nobarfrac{a_1}{-17}~~~\nobarfrac{a_2}{-17}~~~\nobarfrac{a_3}{42}\right\}.
\end{equation*}
The same idea will apply to  
any $(A,o)$.

By monotonicity $\Phi(a_1 \mid A,o)=\Phi(a_2 \mid A,o)$. We will demonstrate that  $\Phi(a_2 \mid A,o)$ and $\Phi(a_3 \mid A,o)$ satisfy a certain identity. Consider the product of $(A,o)$ with the $n$-fold product of $(B,r)$: 
\begin{equation}\label{eq_n_fold_menu}
(A,o) \otimes \underbrace{(B,r)\otimes(B,r)\otimes \cdots  \otimes (B,r)}_{\text{$n$ times}},
\end{equation}
where $n=o(a_3)-o(a_2)=59$. In this menu, the two actions $(a_3,b_0,b_0,\ldots, b_0)$ and $(a_2,b_1,b_1,\ldots, b_1)$ have the same outcome, and thus have the same probability by monotonicity. Therefore, decomposability implies  
\begin{align}
\label{eq_59}
\Phi(a_3 \mid A,o)\cdot \Phi(b_0 \mid B,r)^{59}=\Phi(a_2 \mid A,o)\cdot \Phi(b_1 \mid B,r)^{59}.
\end{align}
Combined with the identities $\Phi(a_1 \mid A,o)=\Phi(a_2 \mid A,o)$ and $\Phi(a_1 \mid A,o)+\Phi(a_2 \mid A,o)+\Phi(a_3 \mid A,o)=1$, this equation pins down $\Phi(A,o)$, which is therefore determined by $\Phi(B,r)$. Since $|B|=2$, we can always choose $\beta \geq 0$ such that $\Phi(B,r)=\mnl^\beta(B,r)$. Since multinomial logit also satisfies the same identities, we conclude that $\Phi(A,o) = \mnl^\beta(A,o)$.

\section{Probit Rules}\label{sec:separable}

Probit rules are a natural and widely used family of stochastic choice rules. In this section we study two natural variants of these rules and explore their relation to our axioms. For simplicity, both are defined for outcomes $\cO=\R$.

\emph{One-shot linear probit} is an independent additive random utility (IARU) model with normal shocks, so that 
\begin{align*}
    \probit(a \mid A,o) = \bP\Big[o(a)+\varepsilon_a=\max_{b\in A} o(b)+\varepsilon_b\Big],
\end{align*}
where $\varepsilon_b$ are independent Gaussians. \emph{Separable linear probit} is a separable IARU model in which each component in a product menu receives its own independent normal shocks  (see the formal definition below). Our main result implies that these rules must violate our axioms. We show that the popular one-shot probit rule violates IID. Perhaps more surprisingly, we demonstrate that separable probit---which does satisfy IID---violates monotonicity. We also show that
probit rules may exhibit negative correlation, unlike random coefficients logit.

Consider the simple one-shot probit rule with 
standard Gaussian  $\mathrm{N}(0,1)$ shocks. We examine its predictions for the menu
\begin{align}\label{eq_binary_menu}
  (B,r) &=
    \left\{\nobarfrac{b_0}{0}~~~\nobarfrac{b_1}{1}\right\}
\end{align}
and its ``square''
\begin{align*}
 (C,s)  = (B,r)\otimes(B,r) =\begin{array}{c|cc}
   & b_0 & b_1 \\
\hline
b_0 & 0 & 1 \\
b_1 & 1 & 2 \\
\end{array}.
    \end{align*}

The probit predictions are  
\begin{align*}
\probit(B,r)\simeq \begin{array}{cc}
    b_0 & b_1 \\
\hline
  0.24 & 0.76 
\end{array} \quad\text{and}\quad \probit(C,s) \simeq \begin{array}{c|cc}
   & b_0 & b_1 \\
\hline
b_0 & 0.033 & 0.175 \\
b_1 & 0.175 & 0.617 \\
\end{array}.
    \end{align*}
Since the marginal probability of choosing $b_1$ from $(C,s)$ is $0.175+0.617=0.792$, which is greater than the choice probability of $b_1$ from $(B,r)$, the probit rule violates~IID. 
We also note that, unlike multinomial logit, which exhibits zero correlation in product menus, the probit predictions for $(C,s)$ exhibit negative correlation, as 
\[
\probit\left((b_1,b_1)\mid C,s\right)\simeq0.617 < 0.627 \simeq \probit\left(b_1 \mid C,s\right)^2. \] 
The reason underlying both effects is that the Gaussian distribution’s tails are too light. When outcome differences are large, the light tails cause probit weights to drop too quickly, much more so than in the logit case. When the differences are small relative to the noise variance, probit is less sensitive to those differences than logit. This leads to insufficient mass on $(b_0,b_0)$ and excessive probability on mixed pairs like $(b_1,b_0)$ and $(b_0,b_1)$, thus generating negative correlation and IID violation.

Separable linear probit ($\sprobit$) is a separable IARU model, where the utility is linear and shocks are Gaussian. As in \eqref{eq:siaru}, it coincides with one-shot linear probit for non-product menus, and for product menus it is given by
\begin{align*}
    \sprobit((a_1,a_2)\mid A,o) 
    &= \sprobit(a_1\mid A,o)\cdot \sprobit(a_2 \mid A,o).
\end{align*}
Unlike one-shot linear probit, in separable linear probit shocks are applied independently in each dimension. Since utilities are linear, we can think of the decision maker as choosing separately in each dimension, or choosing the best combined choice; both yield the same choice probabilities.

By construction, separable linear probit  satisfies  decomposability and thus IID.  We now show that it violates monotonicity.  
Consider the menus
\begin{align*}
    (A_1,o_1) = \left\{\nobarfrac{\text{a}}{0}~~~\nobarfrac{\text{b}}{9}\right\},~~~(A_2,o_2) = \left\{\nobarfrac{\text{c}}{0}~~~\nobarfrac{\text{d}}{6}\right\},~~~(A_3,o_3) = \left\{\nobarfrac{\text{e}}{0}~~~\nobarfrac{\text{f}}{6}\right\},
\end{align*}
and let
\begin{align*}
    (A,o) = (A_1,o_1) \otimes (A_2, o_2) \otimes (A_3,o_3).
\end{align*}
Let $G$ denote the CDF of the difference of two independent standard normals, i.e., a normal distribution with mean zero and variance $2$.
Then $\sprobit(b \mid A_1,o_1) = G(9)$ and $\sprobit(d \mid A_2,o_2) = \sprobit(f \mid A_3,o_3) =  G(6)$.
Since $\sprobit$ satisfies decomposability,
\begin{align*}
  \sprobit ((a,d,f) \mid A,o)&= (1-G(9))\cdot G(6)^2\\
  &< G(9)\cdot (1-G(6))^2\\
  &=\sprobit((b,c,e) \mid A,o).
\end{align*} 
This violates monotonicity, since $(a,d,f)$ has a higher outcome $o(a,d,f) = o_1(a)+o_2(d)+o_3(f) = 12$, while $o(b,c,e) = 9$. 
This again traces back to the light tails of the Gaussian: the probit weights decrease too sharply, leading to a reversal of monotonicity when the rule is extended to product menus in a decomposable way.

The results above demonstrate that widely used rules, such as probit, violate our axioms even in very simple menus.

\section{Utility Representations for Abstract Outcome Spaces $\cO$}\label{sec_general}

So far, we have focused on the outcome spaces $\cO = \R$ and more generally $\cO = \R^n$, and have studied IID and decomposability with respect to addition on these spaces. In the case of $\cO = \R$, this is a direct assumption on how payoffs affect behavior. In this section we take a more abstract approach that avoids using payoffs as a primitive, elucidating the content of our assumptions. For simplicity, we focus in this section on decomposable rules.

Let $\cO$ be a metrizable topological space of outcomes, endowed with a partial order~$\preceq$. One example is, of course, $\R^n$ with the usual partial order. Another example is $\cO=\{\mbox{bounded continuous}\  f\colon \R_{\geq 0}\to \R\}$, which can represent a decision-maker who cares about infinite payoff streams and compares them pointwise. 
The notions of a menu $(A,o)$, a collection of menus $\cM$, and a stochastic choice rule $\Phi$ extend straightforwardly, as do
continuity and monotonicity (with respect to $\preceq$).

To define product menus, we endow the outcome space $\cO$ with a binary operation~$*$ corresponding to combining outcomes in a separable way.
To build intuition, consider  $\cO=\R\times \R_{\geq 0}$, where the first component is interpreted as the mean and the second component as the standard deviation of a stochastic Gaussian monetary reward. 
In this setting, product menus correspond to combined choices in which the stochastic payoffs are independent of each other, so that the operation $*$ is given by
\begin{equation}\label{eq_operation_mean_deviation}
   \left(\nobarfrac{m_1}{\sigma_1}\right)*\left(\nobarfrac{m_2}{\sigma_2}\right)=\left(\nobarfrac{m_1+m_2}{\sqrt{\sigma_1^2+\sigma_2^2}}\right). 
\end{equation}
This structure of outcome space $\cO$ implies that choice probabilities are determined by the expectation and standard deviation. The IID axiom then implies that choice probabilities are not affected by additional decisions that yield independent outcomes. This captures a common assumption reflecting invariance to wealth effects or background risk.

Formally, given two menus $(A_1,o_1)$ and $(A_2,o_2)$, we define their product $(A,o)=(A_1,o_1) \otimes (A_2,o_2)$ by
$$A=A_1 \times A_2,\quad  \text{and}\quad o(a_1,a_2) = o_1(a_1)*o_2(a_2).$$
A rule $\Phi$ is decomposable if 
 \begin{align*}
     \Phi\big((a_1,a_2)\mid A,o \big) = \Phi(a_1 \mid A_1,o_1) \cdot \Phi(a_2 \mid A_2,o_2)
 \end{align*}
 for all $(A_1,o_1),(A_2,o_2)\in \cM$.  Decomposability is a separability assumption of independent behavior in product menus.

The only requirement that we impose on~$*$ is the existence of an irrelevant outcome: there exists $e\in \cO$ such that $e*x=x*e=x$ for any $x\in \cO$. The decision-maker does not care about an action $a$ having outcome $e$ in the sense that combining $a$ with any other action $b$ with outcome~$x$ does not change the decision-maker's perception of $b$.   The operation $*$ defined in~\eqref{eq_operation_mean_deviation}  admits the irrelevant outcome $(0,0)$. It is also commutative and associative, but in general we assume neither commutativity nor associativity. 

For simplicity, we will make an additional assumption of positivity in this section:
\begin{axiom}
    \label{ax:pos}
    A rule $\Phi$ is \textbf{positive} if for any menu $(A,o)$ in $\cM$ and any $a \in A$ it holds that  $\Phi(a \mid A,o) > 0$.
\end{axiom}
The question that we ask in this general setting is: what are the monotone, continuous, positive, decomposable rules?

\medskip

Before stating our main result of this section we will need an additional definition. A function $u\colon \cO\to \R$ is called a \emph{utility representation} of  $(\cO,*)$ if  
\begin{equation}\label{eq_Cauchy}
u(x*y)=u(x)+u(y)\quad \text{for all} \ \  x,y\in \cO.
\end{equation}
In other words, a utility representation assigns a numerical value to each outcome so that combining outcomes in a separable way corresponds to summing their utilities. 
We define monotonicity and continuity of a utility representation with respect to the topology and partial order on the outcome space.

For example, when $(\cO,*)=(\R^n,+)$, any linear function $u(x)=\beta\cdot x$ is a utility representation. Moreover, all continuous utility representations take this form, which follows from a fact about the Cauchy functional equation $u(x)+u(y)=u(x+y)$.\footnote{For $(\cO,*)=(\R,+)$, this dates back to Cauchy; for a general Banach space, see, e.g., \cite{kuczma2009introduction}.} Monotonicity then pins down the signs of the coefficients $\beta_i$: under the standard partial order, it requires $\beta_i\geq 0$. Similarly, when $(\cO,*)$ is the Gaussian outcome space~\eqref{eq_operation_mean_deviation}, the function $u(m,\sigma)=\beta m + \gamma \sigma^2$ is a utility representation. This functional form again exhausts all continuous representations, and the natural monotonicity requirement, which reflects that lower-variance outcomes are chosen more frequently, yields $\gamma\le 0$.

Our usage of the term utility representation for functions $u \colon \mathcal{O} \to \R$ satisfying \eqref{eq_Cauchy} can be motivated by the well-established connection between cardinal utility values and separability: since we think of  $x*y$ as a separable combination of outcomes~$x$ and $y$, the identity~\eqref{eq_Cauchy} is the respective separability condition on utility~$u$. This interpretation relates our analysis 
to the characterization of separable utility by \cite{debreu1959topological}, who shows that choice independence across dimensions pins down a separable utility (uniquely up to affine transformations).

In a similar vein, our second main result relates decomposability and utility representations.
\begin{theorem}
\label{thm:general}
Let $\Phi$ satisfy monotonicity, continuity, positivity and decomposability for $(\cO,*,\preceq)$. Then there exists a continuous, monotone utility representation $u$ of $(\cO,*)$ such that
\begin{align}\label{eq_general_mnl}
    \Phi(a \mid A,o) = \frac{\exp\Big({u\big(o(a)\big)}\Big)}{\sum_{b \in A}\exp\Big({u\big(o(b)\big)}\Big)}
\end{align}
for any menu $(A,o)\in\cM$. 
\end{theorem}

Informally, Theorem~\ref{thm:general} says that there is a canonical way to assign utilities to elements of the outcome space so that choices are governed by a multinomial logit rule with respect to these utilities; moreover, summing these utilities corresponds to combining outcomes in a separable way.
The logit scale parameter is normalized to one, as any positive scale factor can always be absorbed by $u$. 

Theorem~\ref{thm:general} is a generalization of Proposition~\ref{prop:mnl}, which shows a similar result for the case of $(\cO,*) = (\R,+)$. Furthermore, for $(\cO,*) = (\R^n,+)$, all continuous utility representations are of the form $\beta \cdot x$, and so Theorem~\ref{thm:general} is aligned with Theorem~\ref{thm:main} and Proposition~\ref{prop:mnl}.  
The conceptual conclusion we wish to draw from Theorem~\ref{thm:general} is that our main results, characterizing linear logit-type models, hinge on the separability properties of the decomposability and IID axioms, and do not require strong assumptions such as the additive structure of outcomes in $\R^n$.

\section{Proof of Theorem~\ref{thm:main} for $\cO=\R$}\label{sec:thm_main}

The remainder of this paper is a proof of our main theorem, which we hope is of technical interest in its own right. We prove the case of $\cO=\R$ here and extend to $\R^n$ in Appendix~\ref{app:th_multi}. We first describe a classical result of De Finetti regarding partially exchangeable processes. Then, using IID, we extend our stochastic choice rules to infinite products of menus, to which we apply De Finetti's Theorem. Monotonicity ensures the partial exchangeability of a process defined by the extended choice rule. The mixture of i.i.d.\ conclusion of De Finetti's theorem then implies that the choice rule is a mixture of multinomial logit rules.

\subsection{De Finetti's Theorem for Partially Exchangeable Processes}

An important tool in our proof is the De Finetti Theorem for partially exchangeable processes. Let $(X_1,X_2,\ldots)$ be a sequence of random variables, each taking values in some finite set. We denote by $\mathcal{T}$ the tail sigma-algebra of this sequence, i.e., the collection of all events that depend only on the values of $X_i$ for large enough~$i$ and are unaffected by 
modifications to any finite prefix. Let $(X_{i_1},X_{i_2},\ldots)$ be a subsequence. We say that this subsequence is exchangeable if its joint distribution is invariant to any finite permutation of the coordinates.  

Suppose that $\N =\{1,2,\ldots\}$ can be written as a disjoint union of infinite sets $\N = N_0 \cup N_1 \cup N_2 \cup \cdots$. We say that $(X_1,X_2,\ldots)$ is partially exchangeable with respect to this partition if its law is invariant under every finite permutation keeping each $N_k$ intact.

The following is a classical result due to \citet{de1980condition}. It is a generalization of the well-known De Finetti Theorem for exchangeable processes.
\begin{theorem}[De Finetti]
\label{thm:definetti}
 Let $(X_1,X_2,\ldots)$ be a partially exchangeable process, witnessed by the partition $\N = N_0 \cup N_1 \cup N_2 \cup \cdots$. Then for each $k$ there exists a tail-measurable random variable $F_k$ such that, conditioned on the tail sigma-algebra, it holds that (i) the random variables $(X_1,X_2,\ldots)$ are independent, and (ii) for $i \in N_k$ the random variable $X_i$ has distribution $F_k$.
\end{theorem}
The random distributions $F_k$ are the empirical measures. That is, if $\{i_1,i_2,\ldots\}$ is an enumeration of $N_k$, then 
\begin{align*}
    F_k(x) = \lim_n \frac{1}{n}\sum_{m=1}^n 1_{X_{i_m}=x}.
\end{align*}

\subsection{Extending $\Phi$ to Infinite Products}
Let $\cM_{X}$ denote the subset of menus with outcomes in $X \subseteq \R$ and let $\cM=\cM_\R$. Since $\cM$ is closed under $\otimes$, a stochastic choice rule $\Phi$ is defined for any finite product of menus. The following proposition shows that there is a unique way to extend $\Phi$ to countable products of menus taking the form $(A_1,u_1)\otimes (A_2,u_2)\otimes \cdots$. We denote the set of all countable products of menus by $\cM^\infty$.\footnote{Formally, we may identify $\cM^\infty$ with the set of sequences in $\cM$. We only refer to infinite products of menus for notational convenience.}

The next proposition shows that when a rule satisfies monotonicity and IID, we can extend it to countable products of menus. Moreover, this extension satisfies partial exchangeability.

Let $(A_1,o_1),(A_2,o_2),\ldots$ be a sequence of menus, and let $M = (A_1,o_1)\otimes (A_2,o_2)\otimes \cdots \in \cM^\infty$ be their product. Denote by $\Omega = \prod_{i=1}^\infty A_i$ the set of sequences $(a_1,a_2,\ldots)$ corresponding to a choice in each of the menus. For a finite product $(A,o) = (A_1,o_1)\otimes  \cdots \otimes (A_n,o_n)$, a probability measure  $\Phi(A,o)$ over $\prod_{i=1}^n A_i$ describes the probability of each choice. We will extend $\Phi$ to assign to the infinite product menu $M$ a probability distribution over $\Omega$. Given such a measure, denote by $(X_1,X_2,\ldots)$ the random variables corresponding to the choice in each sub-menu, that is, $X_i(a_1,a_2,\ldots) = a_i$. 
\begin{proposition}\label{prop_inf_menus}
    If $\Phi$ satisfies monotonicity and IID, then there is a unique $\Psi$ defined on $\cM^\infty$ such that for every $M= (A_1,o_1)\otimes (A_2,o_2)\otimes \cdots \in \cM^\infty$, $\Psi(M)$ is a probability measure on $\Omega = \prod_{i=1}^\infty A_i$ satisfying
    \[
    \Psi(M)\Bigg(\Big\{a \in  \Omega\,\Big\vert \, a_1 = b_1, a_2 = b_2, \dots, a_n=b_n\Big\}\Bigg) = \Phi\left(\big(b_1,b_2,\dots,b_n\big) \,\middle\vert\,\bigotimes_{i=1}^n(A_i,o_i)\right )
    \]
    for all $n$ and all $(b_1,b_2,\dots,b_n)$. Moreover, if $\N$ is partitioned into classes on which the menus $(A_i,o_i)$ are identical, then  $(X_1,X_2,\ldots)$ is partially exchangeable with respect to this partition.    
\end{proposition}
\begin{proof}
    Equip $\Omega = \prod_{i=1}^\infty A_i$ with the product topology, under which it is compact. Note that the sets of the form $B_{b_1,\ldots,b_n} = \left\{a \in  \Omega \,\middle\vert\, a_1 = b_1, a_2 = b_2, \dots, a_n=b_n\right\}$ are clopen. In particular, given $n \geq 1$ and a probability measure $\mu_n$ on $\prod_{i=1}^n A_i$, denote by $\mathcal{P}(\mu_n)$ the set of probability measures on $\Omega$ that agree with $\mu_n$ on sets of the form $B_{b_1,\ldots,b_n}$:
    \begin{align*}
        \mathcal{P}(\mu_n) = \Big\{\mu\,:\,\mu(B_{b_1,\ldots,b_n}) = \mu_n(\{(b_1,\ldots,b_n)\})\Big\}.
    \end{align*}
    Then $\mathcal{P}(\mu_n)$ is a compact subset of the probability measures on $\Omega$. It is also easily seen to be nonempty. 

    Suppose that $\Phi$ satisfies IID. Denote by $\mu_n$ the measure on $\prod_{i=1}^n A_i$ given by $\Phi(\otimes_{i=1}^n(A_i,o_i))$. By IID, $\mu_{n+1}$ agrees with $\mu_n$ on $B_{b_1,\ldots,b_n}$, so that $\mathcal{P}(\mu_{n+1}) \subseteq \mathcal{P}(\mu_n)$. Since these sets are compact and non-empty, their intersection is non-empty, and so there exists a probability measure $\mu$ on $\Omega$ that agrees with $\mu_n$ on $B_{b_1,\ldots,b_n}$. Since the latter sets generate the sigma-algebra, the measure $\mu$ is unique: $\{\mu\} = \cap_i \mathcal{P}(\mu_n)$. 

Finally, suppose that $\N$ is partitioned as $\N = N_0 \cup N_1 \cup N_2 \cup \cdots$, and that for each block $N_k$, $(A_{i_1}, o_{i_1}) = (A_{i_2}, o_{i_2}) = \cdots$ for indices $i_1, i_2, \ldots$ from~$N_k$. We want to show that the law of $(X_1, X_2, \ldots)$ is invariant under every finite permutation that maps each block $N_k$ into itself. 
   To this end, it suffices to show, without loss of generality, that if we permute $X_{i_1}$ and $X_{i_2}$ then the joint distribution of any long enough prefix $(X_1,X_2,\ldots,X_n)$ of the entire sequence remains unchanged. This follows from the monotonicity of $\Phi$, since the joint distribution of $(X_1,X_2,\ldots,X_n)$ is given by $\Phi(\otimes_{i=1}^n (A_i,o_i))$, and by monotonicity, this distribution assigns equal probabilities to sequences yielding the same total payoff, and payoffs are preserved by permuting copies of the same menu, such as $X_{i_1}$ and $X_{i_2}$, as long as~$n$ is larger than both $i_1$ and $i_2$.
\end{proof}

\subsection{Random Coefficients Logit from De Finetti}

The next proposition is the heart of the proof of Theorem~\ref{thm:main}. Recall that the linear multinomial logit rule is given by
\begin{align*}
    \mnl^\beta(a \mid A,o) = \frac{\exp({\beta \cdot o(a)})}{\sum_{b \in A}\exp({\beta\cdot  o(b)})}
\end{align*}
for $\beta \in \R$. We extend this definition to $\beta \in \{-\infty,+\infty\}$ by letting $\mnl^{+\infty}$ be the rule in which the decision maker chooses uniformly at random one of the actions with the highest outcomes. Likewise, $\mnl^{-\infty}$ is the rule in which the decision maker chooses uniformly at random one of the actions with the lowest outcomes.

We accordingly extend random coefficients logit to allow the random parameter $\beta$ to be chosen from a distribution $\mu$ over the extended reals.
\begin{proposition}\label{lm_Q_mixed logit}
    If $\Phi$ satisfies monotonicity and IID, then there is a random coefficients logit rule $\ml$ with $\mu$ supported on $[0,\infty]$ such that $\Phi|_{\cM_\Z} = \ml|_{\cM_\Z}$.
\end{proposition}

\begin{proof}
Let $(B_1,u_1),(B_2,u_2),\ldots$ be an enumeration of $\mathcal{M}_{\Z}$, the set of all the menus with integer outcomes. Note that $\mathcal{M}_{\Z}$ is countable, because the set of actions $\mathcal{A}$ is countable. We choose  
\begin{align*}
    (B_0,u_0) = \left\{\nobarfrac{\ell}{0}~~~\nobarfrac{h}{1}\right\}.
\end{align*}

Define the sequence of menus $(A_1,o_1),(A_2,o_2),\ldots$ as follows. Write the natural numbers $\N = N_0 \cup N_1 \cup N_2 \cup \cdots$ as a disjoint union of infinite sets, and given $i \in N_k$, set $(A_i,o_i) = (B_k,u_k)$. Hence, there are infinitely many copies of each $(B_k,u_k)$ in the sequence $(A_i,o_i)_i$.

By Proposition~\ref{prop_inf_menus}, there is a unique $\Psi$ defined on $\cM^\infty$ that marginalizes to $\Phi$ on each finite product of menus. Recall that $\Omega = \prod_{i=1}^\infty A_i$. Let $\mP$ denote the probability measure on $\Omega$ given by $\Psi(\bigotimes_{i=1}^\infty (A_i,o_i))$, and let $X_n$ be the coordinate projections, i.e., $X_n(a_1,a_2,\dots)=a_n$, with the tail $\sigma$-algebra of $(X_1,X_2,\dots)$ denoted by $\cT$.

Fix any $k \geq 1$. Enumerate $N_k = \{i_1,i_2,\ldots\}$ and let
\begin{align*}
    (Y_1,Y_2,Y_3,\ldots) = (X_{i_1},X_{i_2},X_{i_3},\ldots).
\end{align*}
Enumerate $N_0 = \{j_1,j_2,\ldots\}$ and let 
\begin{align*}
    (Z_1,Z_2,Z_3,\ldots) = (X_{j_1},X_{j_2},X_{j_3},\ldots).
\end{align*}
Hence $Y_i$ corresponds to the choice in the $i$th copy of $B_k$ and $Z_i$ to the $i$th copy of~$B_0$.

By Proposition~\ref{prop_inf_menus}, the entire process $(X_1,X_2,\ldots)$ is partially exchangeable with respect to the partition  $\N = N_0 \cup N_1 \cup N_2 \cup \cdots$.
It therefore follows by De Finetti  (Theorem~\ref{thm:definetti}) that there are tail-measurable random distributions $F$ and $G$, where $F$ is a distribution over $B_k$ and $G$ is a distribution over $B_0=\{\ell,h\}$, and such that conditioned on the tail we have that $(X_1,X_2,\ldots)$ are independent, with $Y_i$ chosen from $F$, and $Z_i$ chosen from $G$, so that
\begin{align}
    \label{eq:definetti-F-G}
    F(b_k) = \Pr{Y_i=b_k}{\cT}\quad\text{and}\quad G(b_0) = \Pr{Z_i=b_0}{\cT},\quad  \text{for any }i.
\end{align}
The distributions $F$ and $G$ are therefore the (random) empirical distributions of the actions, i.e., 
\begin{align*}
    F(b_k) = \lim_n \frac{1}{n}\sum_{i=1}^n 1_{Y_{i}=b_k}~~\text{ and }~~G(b_0) = \lim_n \frac{1}{n}\sum_{i=1}^n 1_{Z_{i}=b_0}.
\end{align*}
Denote 
\begin{align*}
    \beta = \log \frac{G(h)}{G(\ell)},
\end{align*}
and note that $\beta$ is a random variable taking values in $[-\infty,+\infty]$. Let $\mu$ be the distribution of $\beta$. We claim that $\Phi(B_k,u_k)$ is equal to $\ml(B_k,u_k)$. Since $k$ is arbitrary, and since $(B_k,u_k)_k$ enumerate $\mathcal{M}_{\Z}$, showing this will complete the proof.

Choose any $a,a'\in B_k$ such that $u_k(a) \geq u_k(a')$ and let  $d=u_k(a)-u_k(a')$. By~\eqref{eq:definetti-F-G},
\begin{align*}
    \E{F(a)\cdot G(\ell)^d}&=\E{\Pr{Y_1=a}{\cT}\cdot \Pr{Z_1=\ell}{\cT}\cdots \Pr{Z_{d}=\ell}{\cT}}.
\end{align*}
Since $Y_i$ and $Z_i$ are independent conditioned on the tail, 
\begin{align*}
    \E{F(a)\cdot G(\ell)^d}=\E{\Pr{Y_1=a,Z_1=\ell,\ldots,Z_{d}=\ell}{\cT}},
\end{align*}
and so by the law of total expectation
\begin{align*}
    \E{F(a)\cdot G(\ell)^d}=\Pr{Y_1=a,Z_1=\ell,\ldots,Z_{d}=\ell}.
\end{align*}
Note that the probability on the right hand side is the probability, under $\Psi$, of choosing $a$ in the first copy of $(B_k,u_k)$, and choosing $\ell$ in the first $d$ copies of $(B_0,u_0)$. Since $\Psi$ agrees with $\Phi$ on finite products, and since $\Phi$ is monotone, it follows that this probability is invariant to changing the choices to another set that yields the same total payoff. Hence, since $d = u_k(a)-u_k(a')$, 
\begin{align*}
    \E{F(a)\cdot G(\ell)^d}=\Pr{Y_1=a',Z_1=h,\ldots,Z_{d}=h}.
\end{align*}
By the same argument used above, we have that the right hand side is equal to $\E{F(a')\cdot G(h)^d}$, and so we have shown that 
\begin{align*}
    \E{F(a)\cdot G(\ell)^d}=\E{F(a')\cdot G(h)^d}.
\end{align*}
We will need to show a stronger version of this equality. In particular, let $P$ be monomial in $F(a),F(a'),G(\ell),G(h)$. Then we claim that 
\begin{align}
    \label{eq:F-G-P}
    \E{F(a)\cdot G(\ell)^d\cdot P}=\E{F(a')\cdot G(h)^d\cdot P}.
\end{align}
For example, when $P = F(a')G(\ell)$, then, following the argument above, 
\begin{align*}
    \E{F(a)\cdot G(\ell)^d \cdot P}&=\Pr{Y_1=a,Z_1=\ell,\ldots,Z_{d}=\ell,Y_2=a',Z_{d+1}=\ell}\\
    &=\Pr{Y_1=a',Z_1=h,\ldots,Z_{d}=h,Y_2=a',Z_{d+1}=\ell}\\
    &= \E{F(a')\cdot G(h)^d \cdot P}.
\end{align*}
The general case follows the same idea, introducing an event of the form $Y_i=a$, $Y_i=a'$, $Z_i=\ell$ or $Z_i=h$ for each term in the monomial, using distinct indices $i$ each time. 

By the linearity of expectation, we have that \eqref{eq:F-G-P} holds for any polynomial $P$. Thus, taking $P = F(a)G(\ell)^d-F(a')G(h)^d$, we have that $\E{P^2}=0$, so that 
\begin{align}
  \label{eq:FGd}    F(a)G(\ell)^d=F(a')G(h)^d~~\text{almost surely.}
\end{align}
Note that if $d=0$ then this proof yields that $F(a)=F(a')$ almost surely. Otherwise, $d = u_k(a)-u_k(a')>0$. Let $E_h$ be the event that $G(h) = 0$, which is the event that $\beta = -\infty$. It follows from \eqref{eq:FGd} that $F(a)=0$ on $E_h$. Since $a$ and $a'$ are an arbitrary pair such that $u_k(a)>u_k(a')$, we have $F(b)=0$ for any $b$ that does not yield lowest payoff. Likewise, we have $F(c)=0$ for any $c$ that does not yield the highest payoff conditioned on the event $E_\ell$ where $G(\ell)=0$ and $\beta=+\infty$. We thus have that conditioned on $\beta=+\infty$, $F = \mnl^{+\infty}(B_k,u_k)$, and likewise conditioned on $\beta=-\infty$, $F = \mnl^{-\infty}(B_k,u_k)$.

Outside the union of $E_h$ and $E_\ell$, $\beta$ is finite, and it follows from \eqref{eq:FGd} that  
\begin{align*}
    \frac{F(a)}{F(a')} = \ee^{\beta(u_k(a)-u_k(a'))} = \frac{\ee^{\beta u_k(a)}}{\ee^{\beta u_k(a')}}.
\end{align*}
Hence, also on the event $(E_\ell \cup E_h)^c$ we have that $F = \mnl^\beta(B_k,u_k)$. We have thus shown that $F = \mnl^\beta(B_k,u_k)$, and so
\begin{align*}
    \Phi(a \mid B_k,u_k) = \Pr{Y_i = a} = \E{F(a)} = \E{\mnl^\beta(a \mid B_k,u_k)} = \ml(a \mid B_k,u_k).
\end{align*}
By Lemma~\ref{lm_non-neg_param} from Appendix~\ref{app_monotonicity}, monotonicity implies that $\mu$ is supported on~$[0,\infty]$.
\end{proof}

\subsection{Final Steps}
In the next proposition, we show that the mixing measure of a random coefficients logit rule is uniquely identified by the restriction to $\cM_{\Z}$. 
\begin{proposition}\label{prop_identification}
Suppose $\rcl^{\nu}|_{\cM_{\Z}}=\rcl^{\nu'}|_{\cM_{\Z}}$ with $\nu$ and $\nu'$ supported on $[0,\infty]$. Then $\nu = \nu'$.
\end{proposition}
\begin{proof}
Given $\gamma>0$, define $N_m=\lfloor e^{\gamma m}\rfloor$ and let $(A_m,o_m^\gamma)$ be the menu with $A_m=\{a_1,\ldots,a_{N_m},b\}$, $o_m^\gamma(a_j)=0$ for $j=1,\ldots,N_m$, and $o_m^\gamma(b)=m$. Then
    \begin{align*}
        \lim_{m} \mnl^\beta(b \mid A_m,o_m^\gamma) = \one_{\{\beta>\gamma\}}+
\frac12\one_{\{\beta=\gamma\}}.
    \end{align*}
Hence, by dominated convergence,
\begin{align*}
        \lim_{m}\rcl^\nu(b \mid A_m,o_m^\gamma)
        = \nu((\gamma,\infty])+\frac{1}{2}\nu(\{\gamma\}).
\end{align*}
The same identity holds for $\nu'$. Since the two rules agree on
$\cM_{\Z}$, we get $
\nu((\gamma,\infty])=\nu'((\gamma,\infty])$
at every $\gamma$ that is a not an atom of either $\nu$ or $\nu'$. Since such $\gamma$'s are dense, the intervals $(\gamma,\infty]$ generate all the Borel sets of $[0,\infty]$. 
Thus $\nu=\nu'$.
\end{proof}

Given Propositions~\ref{lm_Q_mixed logit} and \ref{prop_identification}, we are ready to prove our main theorem.
\begin{proof}[Proof of Theorem~\ref{thm:main} for $\cO=\R$]

    We first show  that $\Phi$ coincides with some $\ml$ on $\cM_\Q$. By Proposition~\ref{lm_Q_mixed logit} there is some $\mu$ such that $\Phi$ restricted to $\mathcal{M}_{\Z}$ coincides with $\ml$.
    
    Define for $k = 1,2,\dots$, the rules $\Phi^k$ by \begin{equation}\label{eq_Phik}
    \Phi^k(A,o)=\Phi\left(A,\frac{1}{k}\cdot o\right).
\end{equation} Note that such rules satisfy monotonicity and IID, since $\Phi$ does.  Hence, by Proposition~\ref{lm_Q_mixed logit} again, each $\Phi^k|_{\cM_\Z}=\rcl^{\mu_k}|_{\cM_\Z}$ for some $\mu_k$. By \eqref{eq_Phik} and Proposition~\ref{prop_identification}, it follows that $\mu$ is equal to the push-forward of $\mu_k$ under the map $\beta\to k\beta$. 

For $(A,o)\in \cM_\Q$, there is a positive integer $k$ such that $(A,k\cdot o)\in \cM_\Z$. Thus, 
\[
\Phi(A,o)=\Phi^k(A, k\cdot o) = \rcl^{\mu_k}(A,k\cdot o)=\ml(A,o).
\]
Thus, $\Phi|_{\cM_\Q}=\ml|_{\cM_\Q}$.  
By Lemma~\ref{lm_non-neg_param}, $\mu$ is supported on the non-negative extended reals. By continuity, $\mu(\{+\infty\})=0$. Indeed, consider menus $(A,o_n)$ where $A=\{a,b\}$, $o_n(a)=0$, and $o_n(b)=\frac{1}{n}$. By continuity and monotonicity $\lim_n \Phi(a\mid A,o_n)=\frac{1}{2}$. If, however, $\mu(\{+\infty\})=\varepsilon>0$, then  $\lim_n\Phi(a\mid A,o_n)\leq \frac{1}{2}(1-\varepsilon)$, violating continuity.

    Fix any menu $(A,o)$ and $a \in A$. For $k =1,2,\ldots$, define $\bar{o}_k \colon A \to \Q$ by $\bar{o}_k(a)=\frac{1}{k}\ceil{k\cdot o(a)}$. Since $\bar{o}_k(a) \to o(a)$, by continuity, we have $\Phi(A,o)=\lim_{k}\Phi(A,\bar o_k)$. Hence, $\Phi$ is uniquely determined by $\Phi|_{\cM_\Q}$. Since $\Phi|_{\cM_\Q} = \ml|_{\cM_\Q}$, and since $\ml$ is continuous, it follows that $\Phi = \ml$.  
\end{proof}

\section{Conclusion}\label{sec_conclusion}

This paper explores a novel approach to stochastic choice. We model the 
behavior of an individual across a rich variety of situations, under 
the key assumption that choices remain consistent when decisions are combined in a separable way.
Though often implicit, this assumption underpins the validity 
of experimental analyses that focus on isolated decision problems. In this paper, the corresponding IID axiom is taken as an assumption. But this assumption can be easily tested in a lab, by providing subjects with different menus yielding carefully selected monetary payoffs, and seeing whether their choice probabilities satisfy \eqref{eq_iid}, the defining property of IID.

While each axiom we impose is quite mild, their combination leads to a strong conclusion: choice behavior must follow a 
random coefficients logit rule with a mixing measure fixed across all contexts.
This conclusion is robust to changes in the outcome space and, plausibly, to other features of the model. While we focus on exact adherence 
to the axioms, real decision-makers only satisfy them approximately. We believe our framework can be 
extended to accommodate such deviations---approximate equality in the definition of IID or approximate inequality in the definition of monotonicity---to yield approximate random coefficients logit. 

In our approach, a choice rule predicts behavior in any decision problem,
modeled as a finite collection of vectors interpreted as observable covariates and associated with action labels. As is standard in full-domain characterizations, the axioms are imposed on the entire domain. The practical counterpart of this conceptual exercise is that the analyst assumes the decision maker could in principle face a rich class of menus, and that behavior across these menus would respect separability. This assumption may be motivated by observations from a limited set of decisions, but complete verification would require observing behavior on the whole domain. Once this assumption is adopted, the analyst is naturally led to the random coefficients logit rule, which can then be estimated using a smaller subset of menus. Whether a variant of our axiomatic approach can deliver a (possibly approximate) characterization on a smaller domain remains a natural question for future
work.

\bibliography{refs}

\appendix

\section{Monotonicity of Random Coefficients Logit}\label{app_monotonicity}

 Recall that an extended random coefficients logit is given by $\ml$ for some measure $\mu$ on the extended reals. The following lemma shows that in extended random coefficients logit the distribution of the random response parameter must be supported on the non-negative extended reals in order to achieve monotonicity.
\begin{lemma}\label{lm_non-neg_param}
    The extended random coefficients logit rule with random parameter $\beta \sim \mu$ satisfies monotonicity if and only if it satisfies monotonicity for menus in $\cM_\Z$ if and only if $\mu([-\infty,0))=0$.
\end{lemma}
\begin{proof}[Proof of Lemma~\ref{lm_non-neg_param}]
     Define, for $k=1,2,\dots$, $(A,o_k)=(\{a,b,c\},(0,1,k))$, and suppose, for the sake of contradiction, that $\mu([-\infty,0))>0$. Then 
\begin{align*}
    \lim_{k}\Phi(A,o_k)(a)&=\frac{1}{3}\mu(0)+\lim_k\int_{[-\infty,0)} \frac{1}{1+e^\beta+e^{\beta\cdot k}}\,\dd\mu(\beta)\\
    &=\frac{1}{3}\mu(0)+\int_{[-\infty,0)} \frac{1}{1+e^\beta}\,\dd\mu(\beta)\\
    &> \frac{1}{3}\mu(0)+\int_{[-\infty,0)} \frac{e^\beta}{1+e^\beta}\,\dd\mu(\beta)=\lim_{k}\Phi(A,o_k)(b),
\end{align*} so monotonicity is violated for large enough $k$. On the other hand, when $\mu([-\infty,0))=0$, it is clear that $\ml$ satisfies monotonicity.

\end{proof}

\section{Proof of Proposition~\ref{prop:pos_cor}}\label{ap_prop1}

\begin{lemma}[See, e.g., \cite*{hardy1952inequalities}]\label{lm:ineq}
    If $f,g\colon \R\to \R$ are strictly increasing functions, $\mu$ is a probability measure,  and $f$, $g$, and $fg$ are $\mu$-integrable, then $\int f g\, \dd\mu \geq \int f \,\dd\mu\cdot \int g \,\dd\mu$, with equality if and only if $\mu$ is a Dirac measure.
\end{lemma}
\begin{comment}
\begin{proof}
Note that
    \begin{align*}
      E_\mu[fg]-E_\mu[f]\cdot\mathbb E_\mu[g] &=\frac{1}{2}\int \int (f(x)-f(y))(g(x)-g(y))\,\dd\mu(x)\,\dd\mu(y).  
    \end{align*}
    Since $f$ and $g$ are strictly increasing, the integrand is non-negative. Thus $E_\mu[fg]-E_\mu[f]\cdot\mathbb E_\mu[g] \geq 0$. Moreover, since the integrand is strictly positive for $x \neq y$, if $\mu$ is not a Dirac measure, $\mathbb E_\mu[fg]> \mathbb E_\mu[f]\cdot\mathbb E_\mu[g]$.
\end{proof}
\end{comment}
\begin{lemma}
    \label{lm:increasing}
    Let $(A,o)$ be a menu, and let 
    \begin{align*}
        f(\beta)=\sum_{a \in A} o(a) \cdot \mnl^\beta(a \mid A,o).
    \end{align*}
    Then $f(\beta)$ is increasing in $\beta$, and strictly increasing if $A$ is non-trivial.
\end{lemma}
\begin{proof}
Let $A=\{a_1,\cdots,a_n\}$ and denote $u_i = o(a_i)$. Let $d(\beta) = \sum_i \ee^{\beta u_i}$. Then
\begin{align*}
    d'(\beta) = \sum_i u_i\cdot\ee^{\beta u_i},\,
    d''(\beta) = \sum_i u_i^2\cdot \ee^{\beta u_i}.
\end{align*}
We can rewrite $f(\beta)$ as \[f(\beta)=\frac{d'(\beta)}{d(\beta)},\] and
\begin{align*}
    f'(\beta)&=-\frac{d'(\beta)^2}{d(\beta)^2}+\frac{d''(
    \beta)}{d(\beta)}=-\frac{1}{d(\beta)^2}(d'(\beta)^2-d(\beta)\cdot d''(\beta)).
\end{align*}
By the Cauchy-Schwarz inequality, we have \begin{align*}
    d'(\beta)^2-d(\beta)\cdot d''(\beta) & = \left(\sum_i u_i\cdot\ee^{\beta u_i}\right)^2-\left( \sum_i \ee^{\beta u_i}\right)\cdot\left(\sum_i u_i^2\cdot \ee^{\beta u_i}\right) \leq 0,
\end{align*} 
where the equality holds when 
$u_i\cdot\sqrt{\ee^{\beta u_i}}= \mathrm{const}\cdot\sqrt{\ee^{\beta u_i}}$ for all $i$, i.e., when $u_i$ itself is a constant.
\end{proof}

\begin{proof}[Proof of Proposition~\ref{prop:pos_cor}]
    Let $(A,o)=(A_1,o_1)\otimes (A_2,o_2)$ be a product of non-trivial menus and let \[f_1(\beta)=\sum_{a \in A_1} o_1(a) \cdot \mnl^\beta(a \mid A_1,o_1)\] and \[f_2(\beta)=\sum_{a \in A_2} o_2(a) \cdot \mnl^\beta(a \mid A_2,o_2).\]

It follows that 
\begin{align*}
\sum_{a \in A_1}&\sum_{b\in A_2} o_1(a)\cdot o_2(b)\cdot \ml((a,b)\mid A,o)\\
&= \sum_{a \in A_1}\sum_{b\in A_2}o_1(a)\cdot o_2(b)\cdot \int\mnl^\beta((a,b) \mid A,o)\,\dd\mu(\beta)\\ 
    &= \int \sum_{a \in A_1}\sum_{b \in A_2} o_1(a)\cdot o_2(b)\cdot \mnl^\beta((a,b) \mid A,o)\,\dd\mu(\beta),
\end{align*}
by a change of the order of summation. By the decomposability of $\mnl^\beta$,
\begin{align*}
    &=\int \sum_{a \in A_1}\sum_{b \in A_2} o_1(a)\cdot o_2(b)\cdot \mnl^\beta(a \mid A_1,o_1) \cdot \mnl^\beta(b \mid A_2,o_2)\,\dd\mu(\beta)\\
    &=\int \sum_{a \in A_1}o_1(a)\cdot \mnl^\beta(a \mid A_1,o_1)\sum_{b \in A_2} o_2(b)\cdot \mnl^\beta(b \mid A_2,o_2)\,\dd\mu(\beta)\\
    &=\int f_1(\beta)f_2(\beta)\,\dd\mu(\beta).
\end{align*}
Since $f_1$ and $f_2$ are strictly increasing in $\beta$ (Lemma~\ref{lm:increasing}), 
\begin{align*}
    &\geq \left(\int f_1(\beta)\,\dd\mu(\beta)\right) \cdot \left(\int f_2(\beta)\,\dd\mu(\beta)\right)\\
    &= \left(\sum_{a \in A_1}o_1(a) \cdot \ml(a \mid A_1,o_1)\right)\cdot \left(\sum_{b \in A_2}o_2(b) \cdot \ml(b \mid A_2,o_2)\right).
\end{align*}
Moreover, by Lemma~\ref{lm:ineq}, the inequality holds with equality if and only if $\mu$ is a Dirac measure.

\end{proof}

\section{Proof of Theorem~\ref{thm:main} for $\cO=\R^n$ with $n\geq 2$}\label{app:th_multi}

\begin{proof}
    Let $\Phi$ be a monotone, continuous rule on $\cM_{\R^n}$ satisfying IID. Let $\cB=\{e_1,\dots,e_n\}$ denote the basis of standard unit vectors of $\R^n$.
For $t = 1,\dots, n$, define the rule $\Lambda_t$ on $\cM_{\R}$ by $\Lambda_t(A,o)=\Phi(A,o\cdot e_t)$, where $o \cdot e_t \colon A \to \R^n$ maps $a$ to $o(a) \cdot e_t$.

Since $\Lambda_t$ satisfies monotonicity, continuity and IID, the one-dimensional case proved in \S\ref{sec:thm_main} gives $\Lambda_t=\rcl^{\nu_t}$ for some $\nu_t$ supported on $\R_{\geq 0}$.

    \medskip 

    Let $(B_{n+1},u_{n+1}),(B_{n+2},u_{n+2}),\ldots$ be an enumeration of $\mathcal{M}_{\Z^n}$, the set of all the menus with outcomes in $\Z^n$. Note that $\mathcal{M}_{\Z^n}$ is countable, because the set of actions $\mathcal{A}$ is countable. For $t=1,\dots,n$ choose  
\begin{align}
    \label{eq:B_t}
    (B_t,u_t) = \left\{\nobarfrac{\ell_t}{0}~~~\nobarfrac{h_t}{e_t}\right\}.
\end{align}
Since $(B_{n+1},u_{n+1}),(B_{n+2},u_{n+2}),\ldots$ is an enumeration, the menus \eqref{eq:B_t} will appear twice in $(B_{1},u_{1}),(B_{2},u_{2}),\ldots$.

Define the sequence of menus $(A_1,o_1),(A_2,o_2),\ldots$ in $\mathcal{M}_{\Z^n}$ as follows. Write the natural numbers $\N = N_1 \cup N_2 \cup \cdots$ as a disjoint union of infinite sets, and given $i \in N_k$, set $(A_i,o_i) = (B_k,u_k)$. Hence, there are infinitely many copies of each $(B_k,u_k)$ in the sequence $(A_i,o_i)_i$. Let
$
    \Omega = \prod_{i=1}^\infty A_i.
$ By Proposition~\ref{prop_inf_menus}, there is a unique $\Psi$ defined on $\cM_{\R^n}^\infty$ that marginalizes to $\Phi$ on each finite product of menus.\footnote{Formally, Proposition~\ref{prop_inf_menus} is stated for $\cM_{\R}$, but the exact same proof applies to $\cM_{\R^n}$.} Let $\mP$ denote the probability measure on $\Omega$ given by $\Psi(\bigotimes_{i=1}^\infty (A_i,o_i))$, and let $X_n$ be the coordinate projections, i.e., $X_n(a_1,a_2,\dots)=a_n$, with the tail $\sigma$-algebra of $(X_1,X_2,\dots)$ denoted by $\cT$.

Fix any $k >n $. Enumerate $N_k = \{i_1,i_2,\ldots\}$ and let
\begin{align*}
    (Y_1,Y_2,Y_3,\ldots) = (X_{i_1},X_{i_2},X_{i_3},\ldots).
\end{align*}
For $t=1,\dots,n$, let $N_t = \{j_1^t,j_2^t,\ldots\}$. Let 
\begin{align*}
    (Z_1^t,Z^t_2,Z^t_3,\ldots) = (X_{j_1^t},X_{j^t_2},X_{j^t_3},\ldots),
\end{align*}
Hence $Y_i$ corresponds to the choice in the $i$th copy of $B_k$ and $Z_i^t$ to the $i$th copy of $B_t$, as defined in \eqref{eq:B_t}.

By Proposition~\ref{prop_inf_menus}, we know that $(Y_1,Y_2,\ldots)$ are exchangeable, as are $(Z_1^t,Z_2^t,\ldots)$. It therefore follows by De Finetti  (Theorem~\ref{thm:definetti}) that there are tail-measurable random distributions $F, G_1,\ldots,G_n$, where $F$ is a distribution over $B_k$ and $G_t$ is a distribution over $B_t=\{\ell_t,h_t\}$, and such that conditioned on the tail we have that $(X_1,X_2,\ldots)$ are independent, with $Y_i$ chosen from $F$, and $Z_i^t$ chosen from $G_t$, so that
\begin{align}
    \label{eq:definetti-F-G_t}
    F(b_k) = \mathbb{P}[Y_i=b_k\,|\,\cT]~~\text{and}~~G_t(b_t) = \mathbb{P}[Z_i^t=b_t\,|\,\cT], \text{ for any }i.
\end{align}
The distributions $F, G_1,\ldots,G_n$ are therefore the (random) empirical distributions of the actions, i.e., 
\begin{align*}
    F(b_k) = \lim_n \frac{1}{n}\sum_{i=1}^n 1_{Y_{i}=b_k}~~\text{ and }~~G_t(b_t) = \lim_n \frac{1}{n}\sum_{i=1}^n 1_{Z_{i}^t=b_t}\quad\text{ a.s.}
\end{align*}

Denote by $\beta$ the $n$-dimensional random variable with coordinates
\begin{align*}
    \beta_t = \log \frac{G_t(h_t)}{G_t(\ell_t)},
\end{align*}
 and note that $\beta$ takes values in $[-\infty,+\infty]^n$. Let $\mu$ be the distribution of $\beta$ and $\mu_t$ be the marginal distribution of $\beta_t$.

We claim that $\Phi(B_k,u_k)$ is equal to $\ml(B_k,u_k)$. 
To this end, choose any $a,a'\in B_k$, let $d_t$ denote the $t$-th coordinate of $u_k(a)-u_k(a')$, and let $d_t^+=\max\{d_t,0\}$ and $d_t^-=\max\{-d_t,0\}$. By \eqref{eq:definetti-F-G_t}, $Y_i$ and $Z_i^1,\dots,Z_i^n$ are independent conditioned on the tail. By the same argument used in the proof of Proposition~\ref{lm_Q_mixed logit} we conclude that  
\begin{align}\label{eq:FGd_n}
    F(a)&\cdot G_1(\ell_1)^{d_1^+}\cdots G_n(\ell_n)^{d_n^+}\cdot G_1(h_1)^{d_1^-}\cdots G_n(h_n)^{d_n^-}\nonumber\\
    &=F(a')\cdot G_1(\ell_1)^{d_1^-}\cdots G_n(\ell_n)^{d_n^-}\cdot G_1(h_1)^{d_1^+}\cdots G_n(h_n)^{d_n^+}~~\text{almost surely.}
\end{align}

Let $\R_t$ denote the one-dimensional vector space spanned by $e_t$. Note that for $(B_k,u_k) \in \cM_{\R_t}$, $d_i^+=d_i^-=0$ for all $i \neq t.$ Hence, in this case we have \[\frac{F(a)}{F(a')}=\ee^{\beta_t d_t}=\frac{\ee^{\beta_t u_k(a)_t}}{\ee^{\beta_t u_k(a')_t}},\]
 and we may conclude that $\Lambda_t|_{\cM_\Z}=\rcl^{\mu_t}|_{\cM_\Z}$. Thus, by Proposition~\ref{prop_identification}, $\mu_t=\nu_t$, so the support of $\mu_t$ is contained in $\R_{\geq 0}$. Thus, $\beta$ is finite (and $F$ is non-degenerate) almost surely. For any $(B_k,u_k)$ it follows from \eqref{eq:FGd_n} that  
\begin{align*}
    \frac{F(a)}{F(a')} = \ee^{\beta \cdot(u_k(a)-u_k(a'))} = \frac{\ee^{\beta \cdot u_k(a)}}{\ee^{\beta \cdot u_k(a')}}.
\end{align*}
We have thus shown that $F = \mnl^\beta(B_k,u_k)$, and so
\begin{align*}
    \Phi(a \mid B_k,u_k) = \mathbb{P}[Y_i = a] = \mathbb{E}[F(a)] = \mathbb{E}[\mnl^\beta(a \mid B_k,u_k)] = \ml(a \mid B_k,u_k).
\end{align*}

We now show that $\Phi|_{\cM_{\Q^n}}=\ml|_{\cM_{\Q^n}}$. As in the proof of the one-dimensional case,
we define, for $k=1,2,\dots$,
$$
\Phi^k(A,o)=\Phi\left(A,\frac{1}{k}\cdot o\right).
$$
The integer-outcome argument above applies to each $\Phi^k$, so
$\Phi^k|_{\cM_{\Z^n}}=\rcl^{\mu_k}|_{\cM_{\Z^n}}$ for some measure
$\mu_k$ on $\R_{\geq 0}^n$. Let $\tilde{\mu}_k$ be the image of $\mu_k$ under the map $\beta\to k\beta$. We claim that $\mu=\tilde{\mu}_k$.

We use the following identification fact generalizing Proposition~\ref{prop_identification}. Let $\nu,\nu'$ be mixing
measures on $\R_{\geq 0}^n$ such that $\rcl^{\nu}|_{\cM_{\Z^n}}=\rcl^{\nu'}|_{\cM_{\Z^n}}$. We claim that $\nu = \nu'$. Fix
nonempty $S\subseteq\{1,\ldots,n\}$ and
$\gamma\in\R_{>0}^S$. For $t\in S$, set
$N_{t,m}=\lfloor e^{\gamma_t m}\rfloor$ and consider the menu with
$N_{t,m}$ actions of outcome $0$ and one action $b_t$ of outcome
$m e_t$. In the product of these menus over $t\in S$, the probability of
choosing all distinguished actions under $\mnl^\beta$ converges to
$$
\prod_{t\in S}\left(
\one_{\{\beta_t>\gamma_t\}}+
\frac12\one_{\{\beta_t=\gamma_t\}}\right).
$$
Thus, whenever the coordinates of $\gamma$ are not atoms of the corresponding one-dimensional marginals,
$$
\nu(\{\beta_t>\gamma_t\ \forall t\in S\})
=
\nu'(\{\beta_t>\gamma_t\ \forall t\in S\}).
$$
Such $\gamma$'s are dense. Hence, the corresponding upper cylinders, together with $\R_{\geq 0}^n$, generate the
Borel sets of $\R_{\geq 0}^n$.\footnote{The sets with $S=\{1,\ldots, n\}$ are enough to generate the Borel sigma-algebra of $\R_{>0}^n$. Smaller~$S$ are needed for the faces of $\R_{\geq 0}^n$, where some coordinates are zero.}

Applying this identification fact with $\nu=\mu$ and
$\nu'=\tilde{\mu}_k$, we obtain $\mu=\tilde{\mu}_k$.
Now consider $(A,o)\in\cM_{\Q^n}$ and choose a positive integer $k$ such that $(A,k\cdot o)\in \cM_{\Z^n}$. Thus, 
\[
\Phi(A,o)=\Phi^k(A, k\cdot o) = \rcl^{\mu_k}(A,k\cdot o)=\ml(A,o).
\]
Thus, $\Phi|_{\cM_{\Q^n}}=\ml|_{\cM_{\Q^n}}$. Fix any menu $(A,o)$ and $a \in A$. For $k=1,2,\ldots$, define $\bar{o}_k \colon A \to \Q^n$ by $\bar{o}_k(a)=\frac{1}{k}\ceil{k\cdot o(a)}$, where the ceiling is applied componentwise.
Since $\bar{o}_k(a) \to o(a)$, by continuity, we have $\Phi(A,o)=\lim_{k}\Phi(A,\bar o_k)$. Hence, $\Phi$ is uniquely determined by $\Phi|_{\cM_{\Q^n}}$. Since $\Phi|_{\cM_{\Q^n}} = \ml|_{\cM_{\Q^n}}$, and since $\ml$ is continuous, it follows that $\Phi = \ml$.  

\end{proof}

	In some natural applications of stochastic choice, outcomes are limited to a subset of Euclidean space, such as the positive orthant. 
    Our characterization still applies, provided the outcome space is rich enough. Indeed, as the following corollary shows, a stochastic choice rule defined for menus with outcomes in a convex cone that satisfies the axioms can be extended to a rule on $\R^n$ that satisfies the axioms.
    \begin{corollary}
        Let $C\subseteq \R^n$ be a full-dimensional convex cone, and let $\Phi$ be a stochastic choice rule on $\cM_{C}$ that satisfies monotonicity, continuity, and IID. Then $\Phi=\ml$ for some $\mu$ supported on $\R_{\geq 0}^n$.
    \end{corollary}
    \begin{proof}
        Let $C$ and $\Phi$ be as in the corollary. Define $\Psi$ on $\cM_{\R^n}$ by $\Psi(A,o)=\Phi(A,o+c)$, where $c\in C$ is such that $o(a)+c\in C$ for all $a\in A$.
        By IID, this definition does not depend on the choice of $c$. To see
this, suppose $c$ and $c'$ are both admissible shifts. By the cone
property, $c+c'$ is also admissible. Applying IID to the product of
$(A,o+c)$ with a singleton menu having outcome $c'$ gives
$\Phi(A,o+c)=\Phi(A,o+c+c')$, and, similarly, applying IID to the
product of $(A,o+c')$ with a singleton menu having outcome $c$ results in
$\Phi(A,o+c')=\Phi(A,o+c+c')$. Hence $\Phi(A,o+c)=\Phi(A,o+c')$, so
$\Psi$ is well-defined.
It is obvious that $\Psi$ is monotone. $\Psi$ is moreover continuous as the same $c$ from the interior of $C$ works for all menus sufficiently close to $(A,o)$.        
        Finally, for $\Psi(A_1,o_1)=\Phi(A_1,o_1+c_1)$ and $\Psi(A_2,o_2)=\Phi(A_2,o_2+c_2)$ and $(A,o)=(A_1,o_1)\otimes(A_2,o_2)$, we have
        \[
        \Psi(a_1\mid A,o)=\Phi(a_1\mid A,o+c_1+c_2)=\Phi(a_1\mid A_1,o_1+c_1)=\Psi(a_1\mid A_1,o_1),
        \]
        for all $a_1\in A_1$. Thus $\Psi$ satisfies IID and is a random coefficients  logit rule. Since $\Psi$ agrees with $\Phi$ on $C$, $\Phi$ is a random coefficients logit rule.
    \end{proof}
\section{Proof of Proposition~\ref{prop:mnl}}\label{app_mnl_proof}
\begin{proof}
Since decomposability implies IID, by Theorem~\ref{thm:main}, $\Phi=\ml$ for some $\mu$ supported on $\R_{\geq 0}^n$. For $k=1, \dots, n$, define the stochastic choice rule $\Phi^k$ for the outcome space $\cO=\R$ by $\Phi^k(a \mid A,o)=\Phi(a \mid A,o_k)$, where $o_k(a)=o(a)\cdot e_k$ where $e_k$ is the standard basis vector whose $k$th component is $1$ and whose other components are $0$. It follows that
\begin{align*}
    \Phi^k(a\mid A,o)&=\int \mnl^\beta(a\mid A,o_k)\,\dd \mu(\beta)\\
    &=\int \mnl^{\beta_k}(a\mid A,o)\,\dd \mu(\beta) \\
    &=\int \mnl^{\beta_k}(a\mid A,o)\,\dd \mu_k(\beta_k),
\end{align*}
where $\mu_k$ is the $k$th marginal of $\mu$. Note that $\Phi^k$ inherits decomposability from $\Phi$. By Proposition~\ref{prop:pos_cor}, $\mu_k$ is a Dirac measure, since decomposability implies that $\Phi^k$ exhibits zero correlation on product menus. Since each marginal of $\mu$ is a Dirac measure, $\mu$ itself is a Dirac measure, i.e., $\Phi=\mnl^\beta$ for some $\beta \in \R_{\geq 0}^n$.
\end{proof}

\section{Proof of Theorem~\ref{thm:general}}\label{app_general_proof}

\begin{proof}
Recall that $e$ denotes the identity element of $(\cO,*)$. 
For each element $x\in \cO$ of the outcome space, fix a  menu 
\begin{align*}
(A_x,o_x) =
\left\{\nobarfrac{a_e}{e}~~~\nobarfrac{a_x}{x}\right\}
\end{align*}
that has two actions with outcomes $e$ and $x$. Let $p_x = \Phi(a_x \mid A_x,o_x)$ be the probability that the action with outcome $x$ is chosen in this menu.

Define the function $u\colon \cO\to \R$ by 
\begin{equation}\label{eq_explicit_formula_u_app}
u(x)=\ln\frac{p_x}{1-p_x}.
\end{equation}
Note that this logarithm is finite by the positivity of $\Phi$.

We now demonstrate that $u$ is a utility representation of $\cO$, i.e., that it satisfies the generalized  Cauchy equation  $u(x*y)=u(x)+u(y)$ for all $x,y\in \cO$. Consider a product menu 
$$(B,o)=\Big((A_x,o_x)\otimes (A_y,o_y)\Big)\otimes (A_{x*y},o_{x*y}).$$
Recall that the associativity of~$*$ is not assumed, so we must be careful about the order of operations. 
The constructed product menu contains actions 
$$b=\big((a_e,a_e),a_{x*y}\big)\qquad\mbox{and}\qquad b'=\big((a_x,a_y),a_e\big).$$ 
Computing their outcomes, we get
$$o(b)= (e*e)*(x*y)=x*y\qquad\mbox{and}\qquad o(b')=(x*y)*e=x*y,$$
where we used the fact that $e$ is both a left and a right identity.
Since the outcomes of $b$ and $b'$ are the same, monotonicity of $\Phi$ implies
$$\Phi(b \mid B,o)=\Phi(b' \mid B,o).$$
By decomposability of $\Phi$, this identity can be rewritten as follows
$$
(1-p_x) \cdot (1-p_y) \cdot p_{x * y} = p_x \cdot p_y \cdot (1-p_{x * y}).
$$
Taking the logarithm and using the definition of $u$, we obtain
$$u(x*y)=u(x)+u(y)$$
and conclude that $u$ is a utility representation of $\cO$. We stress that the proof of this fact uses neither associativity nor commutativity of~$*$.

We now consider an arbitrary menu $(A,o)$ and demonstrate that $\Phi(A,o)$ is given by  multinomial logit with the constructed utility function~\eqref{eq_general_mnl}. Let $a,b\in A$ be two distinct actions. We express the ratio ${\Phi(a \mid A,o)}/{\Phi(b \mid A,o)}$ in terms of the constructed utility representation~$u$. Denote the outcomes by $x=o(a)$ and $y=o(b)$ and 
consider a new menu
$$\Big((A_x,o_x)\otimes(A,o)\Big)\otimes (A_y,o_y).$$
The two actions 
$$\big((a_e,a),a_y\big)\quad\mbox{and}\quad \big((a_x,b),a_e\big)$$
have equal outcomes. Indeed, $(e*x)*y=x*y$ and $(x*y)*e=x*y$ by the assumption that $e$ is both a left and a right identity element. By monotonicity, $\Phi$ assigns equal probabilities to these actions.
Applying decomposability, we obtain 
$$(1-p_x)\cdot \Phi(a \mid A,o)\cdot p_y=p_x\cdot \Phi(b \mid A,o)\cdot (1-p_y). $$
This equality can be rewritten as
$$
\frac{\Phi(a \mid A,o)}{\Phi(b \mid A,o)}=\frac{p_x/(1-p_x)}{p_y/(1-p_y)}
$$
and thus
$$\frac{\Phi(a \mid A,o)}{\Phi(b \mid A,o)}=\frac{ \exp\Big({u\big(o(a)\big)}\Big)}{\exp\Big({u\big(o(b)\big)}\Big)}.$$
Since $a$ and $b$ were arbitrary, and $\sum_{b\in A} \Phi(b \mid A,o)=1$, we conclude that
\begin{equation}\label{eq_general_mnl_appendix}
 \Phi(a \mid A,o) = \frac{\exp\Big({u\big(o(a)\big)}\Big)}{\sum_{b \in A}\exp\Big({u\big(o(b)\big)}\Big)},
 \end{equation}
i.e., $\Phi$ is multinomial logit.

To verify that $u$ is monotone, take  $x\preceq y$ and consider the binary menu with outcomes $x$ and $y$. By monotonicity of $\Phi$, the action with outcome $y$ is chosen with probability at least as large as the action with outcome $x$. Using the representation~\eqref{eq_general_mnl_appendix}, this implies $\exp(u(y))\geq \exp(u(x))$, and hence $u(y)\geq u(x)$.

It remains to check the equivalence between the continuity of $\Phi$ and that of~$u$. If $u\colon \cO\to\R$ is a continuous utility representation, then $\Phi$ given by~\eqref{eq_general_mnl_appendix} is continuous since the right-hand side of~\eqref{eq_general_mnl_appendix} is a continuous function of the profile of outcomes. In the opposite direction, we suppose that $\Phi$ is continuous and show that $u(x_n)\to u(x)$
for any sequence $x_n\to x$ in $\cO$. Fix a binary set of actions $A=\{a,b\}$ and consider outcome functions $o_n$ and $o$ given by
 $o_n(a)=o(a)=x$,  $o_n(b)=x_n$, and $o(b)=x$. Thus $(A,o_n)$ converges to the menu with identical outcomes $(A,o)$. By continuity, $\Phi(A,o_n)$ converges to the uniform distribution~$\Phi(A,o)$. By~\eqref{eq_general_mnl_appendix}, we get 
 $$
 \frac{\exp\Big({u\big(x_n\big)}\Big)}{\exp\Big({u\big(x\big)}\Big)}=\frac{\Phi(b\mid A,o_n)}{\Phi(a \mid A,o_n)}\to \frac{\Phi(b\mid A,o)}{\Phi(a\mid A,o)}=1 
 $$
 and so $u(x_n)$ converges to $u(x)$. Thus $u$ is continuous.

 \end{proof}

\section{Restrictions of Monotonicity and IID}\label{sec:restriction}
The family of random coefficients logit rules is parameterized by a probability measure, which is an infinite dimensional object. This gives this family considerable flexibility to model a wide range of behavior. Nevertheless, knowing that a rule belongs to this family imposes significant restrictions on the choice probabilities, even within simple menus. As we discussed in \S\ref{sec_axiom_implications}, for the menu
\begin{align*}
    (A,o) &= \left\{\nobarfrac{a_1}{0}~~~\nobarfrac{a_2}{2}~~~\nobarfrac{a_3}{7}\right\}
\end{align*}
it holds for every random coefficients logit rule $\Phi$ that  $\Phi(a_1 \mid A,o) = 5\%$ implies that $\Phi(a_3 \mid A,o) \geq 83.8\%$. Indeed, this lower bound of the choice probability of $a_3$ turns out to be $\mnl^\beta(a_3 \mid A,o)$ for the $\beta$ that satisfies $\mnl^\beta(a_1 \mid A,o) = 5\%$. The following proposition shows that this is a general property of random coefficients  logit rules for $\cO  =\R$. 
\begin{proposition}\label{prop_restrictions}
Let $A=\{a_1, \dots, a_n\}$ and $o(a_1) \leq o(a_2) \leq \cdots \leq o(a_n)$ with at least one strict inequality. Let $\beta \in \R$ be the unique parameter satisfying $\mnl^{\beta}(a_1 \mid A,o) = \Phi(a_1 \mid A,o)$. If $\Phi$ is a random coefficients logit rule, then $\Phi(a_n \mid A,o) \geq \mnl^{\beta}(a_n \mid A,o)$.
\end{proposition}
By a symmetric argument, we can get a lower bound on the probability of $a_1$, given the probability of $a_n$: if $\gamma$ is such that $\mnl^{\gamma}(a_n \mid A,o) = \Phi(a_n \mid A,o)$, then $\Phi(a_1 \mid A,o) \geq \mnl^{\gamma}(a_1 \mid A,o)$.

Proposition~\ref{prop_restrictions} highlights a property of random coefficients  logit, independently of our main results. But since our axioms imply random coefficients logit, it follows from this proposition and our main results that every rule that satisfies our axioms also satisfies this property. We do not know a direct proof of this.

Before proving this proposition we state and prove a lemma.

\begin{lemma}\label{lm_risk_coeffs}
Let $L(\beta)=\mnl^\beta(a_1 \mid A,o)$ and $H(\beta)=\mnl^\beta(a_n \mid A,o)$, where $A=\{a_1, \dots, a_n\}$ and $o(a_1) \leq o(a_2) \leq \cdots \leq o(a_n)$, with at least one strict inequality.  Then $-\frac{H''}{H'} \leq -\frac{L''}{L'}$. 
\end{lemma}
\begin{proof}
Because logit probabilities are invariant to adding a common constant to all outcomes, assume without loss of generality that $o(a_1)=0$. Let $o_i$ denote $o(a_i)$, and let $d(\beta)=\sum_i \ee^{\beta o_i}$. Then $0\leq o_i\leq o_n$ for all $i$, with $o_n>0$, and
\begin{align*}
    d'(\beta) &= \sum_i o_i \ee^{\beta o_i} < o_n\cdot d(\beta),\\
    d''(\beta) &= \sum_i o_i^2 \ee^{\beta o_i} \leq o_n\cdot d'(\beta).
\end{align*}
The derivatives of $L$ are given by:
\begin{align*}
    L' &= -\frac{d'}{d^2} >  -\frac{o_n \cdot d}{d^2} = -o_n \cdot L \\
    L'' &= \frac{2(d')^2 - d''\cdot d}{d^3}
\end{align*}
Since $o_1=0$, we have $H= \ee^{\beta o_n}L$. Then
\begin{align*}
    H' &= o_n \ee^{\beta o_n} L + e^{\beta o_n} L' = \ee^{\beta o_n} (o_n L + L') \\
    H'' &= o_n^2 \ee^{\beta o_n} L + 2 o_n \ee^{\beta o_n} L' + \ee^{\beta o_n} L''.
\end{align*}
Thus, we have:
\begin{align*}
    \frac{H''}{H'} - \frac{L''}{L'} &= \frac{o_n^2 L + 2 o_n L' + L''}{o_n L + L'} - \frac{L''}{L'} \\
    &= \frac{o_n^2 L\cdot L' + 2 o_n (L')^2 - o_n L\cdot L''}{L'(o_n L + L')}.
\end{align*}
Since $L' < 0$ and $o_nL+L' > 0$, the denominator of the above expression is negative. Moreover the numerator simplifies to 
\[
\frac{o_n(d''-o_nd')}{d^3} \leq 0,
\]
so the overall expression is non-negative, as desired.
\end{proof}

\begin{proof}[Proof of Proposition~\ref{prop_restrictions}]
         Consider $A=\{a_1, \dots, a_n\}$ and outcomes $o(a_1) \leq o(a_2) \leq \cdots \leq o(a_n)$ with at least one strict inequality.
     Without loss of generality, suppose $o(a_1)=0$. Define $d(\beta)=\sum_i \ee^{\beta\cdot o(a_i)}$. Let $L(\beta) = \frac{1}{d(\beta)}$ and $H(\beta)=\frac{\ee^{\beta \cdot o(a_n)}}{d(\beta)}$ denote the logit choice probabilities of the low and high alternatives, respectively, as a function of $\beta$. 
    
    Denote $p = \Phi(a_1 \mid A,o)$. Since $L(\beta)=1/d(\beta)$ is strictly decreasing, the $\beta$ solving $L(\beta) = p$ is uniquely pinned down by $p$. Let $F(\beta)=-L(\beta)$, so that $H$ and $F$ are increasing functions.

By Lemma~\ref{lm_risk_coeffs},
\[
-\frac{H''}{H'}\leq -\frac{L''}{L'} = -\frac{F''}{F'}.
\]

Hence,  by \cite*{pratt1964risk}, $F\circ H^{-1}$ is increasing and concave, i.e., $L \circ H^{-1}$ is decreasing and convex, and its inverse $H \circ L^{-1}$ is convex. Because $\Phi$ is random coefficients logit, there is a probability measure $\mu$ such that $\Phi(a_n\mid A,o) = \int H \,\dd\mu$ and $p = \int L \,\dd\mu$. For such $\mu$, we have
\begin{align*}
    \Phi(a_n\mid A,o)&=\int H \,\dd\mu = \int H \circ L^{-1} \circ L \,\dd\mu \geq H \circ L^{-1}\left( \int L \,\dd\mu\right) \\
    &= H \circ L^{-1}(p)=H(\beta)=\mnl^{\beta}(a_n \mid A,o).
\end{align*}
 The inequality follows from Jensen's inequality since $H \circ L^{-1}$ is convex. 
 \end{proof}

\end{document}